\documentclass{elsart}

\usepackage{natbib}
\usepackage{graphicx}
\usepackage{amsmath, amssymb}
\usepackage{lscape} % Wes Custom
\usepackage{epstopdf}
\usepackage{longtable}

\begin{document}

%%%%
%Wes' Custom Commands Section
%%%%
\newcommand{\atotal}{\alpha = 0.65 \pm 0.05}   %1-sigma errors

\newcommand{\mtotal}{\ensuremath{m_{o} = 23.42 \pm 0.13}} 
\newcommand{\mtwelvek}{\ensuremath{m_{o,12k} =  23.5 \pm 0.16}}
\newcommand{\mMEGA}{\ensuremath{m_{o,MEGA} = 24.16 \pm 0.11}}
\newcommand{\mCTIOmine}{\ensuremath{m_{o,CTIO2001} = 23.3 \pm 0.16}}
\newcommand{\mCTIOmatts}{\ensuremath{m_{o,CTIO2002} = 23.4 \pm 0.16}}
\newcommand{\mSSU}{\ensuremath{m_{o,SSU} = 23.1 \pm 0.13}}
\newcommand{\mSSN}{\ensuremath{m_{o,SSN} = 23.0 \pm 0.15}}
\newcommand{\mGOO}{\ensuremath{m_{o,G00} = 23.3 \pm 0.18}}
\newcommand{\mAF}{\ensuremath{m_{o,AF} = 23.6 \pm 0.18}}
\newcommand{\mAKL}{\ensuremath{m_{o,AKL} = 23.4 \pm 0.20}}
\newcommand{\mFoneG}{\ensuremath{m_{o,TE1G} = 23.4 \pm 0.15}}
\newcommand{\mFtwoG}{\ensuremath{m_{o,TE2G} = 23.4 \pm 0.18}}
\newcommand{\mFthreeG}{\ensuremath{m_{o,TE3G} = 23.5 \pm 0.16}}
\newcommand{\mFfiveM}{\ensuremath{m_{o,TE5M} = 23.2 \pm 0.21}}
\newcommand{\mOTHER}{\ensuremath{m_{o,OTHER} = 24.70 \pm 0.09}}

\newcommand{\arcmin}{^{'}}

\newcommand{\aap}{A\&AS}           
\newcommand{\aj}{AJ}           
\newcommand{\nat}{Nature}
\newcommand{\apj}{ApJ}
\newcommand{\icarus}{Icarus}
\newcommand{\apjs}{ApJS}
\newcommand{\pasp}{PASP}
\newcommand{\mnras}{MNRAS}
\begin{frontmatter}

\title{The Kuiper Belt Luminosity Function from $m_{R}=21$ to $26$.}

\author[label1,label2]{Wesley C. Fraser}
\ead{wesley.fraser@nrc.ca}
\author[label1,label2,label4,label5]{JJ Kavelaars}
\author[label3,label4,label5]{M. J. Holman}
\author[label1]{C. J. Pritchet}
\author[abc,label4]{B. J. Gladman}
\author[label3]{T. Grav}
\author[label2]{R. L. Jones}
\author[label2]{J. MacWilliams}
\author[label7,label3]{J.-M. Petit}

\address[label1]{Dept. of Physics and Astronomy, University of Victoria, Victoria, BC V8W 3P6, Canada}
\address[label2]{Hertzberg Institute of Astrophysics, National Research Council of Canada, Victoria, BC V9E 2E7, Canada}
\address[label3]{Harvard-Smithsonian Center for Astrophysics, 60 Garden St., MS 51, Cambridge, MA 02138, USA}
\address[label4]{Guest Observer, Canada France Hawaii Telescope}
\address[label5]{Guest Observer, Cerro Tololo Inter-American Observatory}
\address[abc]{Dept. of Physics and Astronomy, University of British Columbia, Vancouver, BC V6T 1Z1, Canada}
\address[label7]{Observatoire de Besan\c{c}on, 25010 Besan\c{c}on, Cedex, France}

Submitted to {\it Icarus}\\
Accepted, February 6, 2008\\

Manuscript Pages: 50\\
Tables: 7\\
Figs: 8

\newpage

\begin{abstract}
We have performed an ecliptic imaging survey of the Kuiper belt with our deepest and widest field achieving a limiting flux of $m(g')_{50\%} \sim 26.4$, with a sky coverage of 3.0 square-degrees. This is the largest coverage of any other
Kuiper belt survey to this depth. We
detect 72 objects, two of which have been
previously observed. We have improved the Bayesian maximum likelihood  fitting technique presented in Gladman et al. (1998) to account for calibration and sky density variations and have used this to determine the luminosity
function of the Kuiper belt. Combining our detections with previous surveys,
we find the luminosity function is well represented by a single power-law
with slope $\alpha = 0.65\pm0.05$ and an on ecliptic sky density of 1 object per square-degree brighter than $m_{R}=23.42 \pm 0.13$. Assuming constant albedos, this slope
suggests a differential size-distribution slope of $4.25\pm0.25$, which is steeper than the
Dohnanyi slope of 3.5 expected if the belt is in a state of
collisional equilibrium. We find no evidence for a roll-over or knee
in the luminosity function and reject such models brightward of $m(R)\sim24.6$.
\end{abstract}

\begin{keyword}
COLLISIONAL PHYSICS
KUIPER BELT
\end{keyword}

\newpage

\end{frontmatter}

% main text
{\bf Proposed Running Head:} Kuiper Belt Luminosity Function.\\

{\bf Editorial Correspondence} \\
Wesley C. Fraser\\
Dept. of Physics and Astronomy\\
University of Victoria\\
Victoria, BC, Canada\\
V8W 3P6

\newpage

\section{Introduction}
The study of extrasolar debris and dust disks has revealed that, for
at least some of these disks to exist as we see them, there must be a
source which is responsible for replenishment of small-grain dust in
the disk. Otherwise, due to radiation pressure, the small grain dust
would be blown out of their stellar systems on time scales shorter
than the age of the star. A disk of planetesimals embedded in the dust
disk which is undergoing collisional evolution is a likely source of
this dust. Disruptive collisions could produce the necessary influx of
dust to extend the lifetime of the disk beyond the dust blow-out
time. See \citet{Meyer2006} for a current review of debris disks.

The Kuiper belt is analogous to these extra-solar planetesimal disks,
and provides an excellent laboratory to study and understand the
properties of these planetesimals and the processes that affect them,
including collisional processes, tensile strengths, compositions, and
the mechanisms by which they formed. Knowledge of the size
distribution in the belt can constrain much of this
information. The size distribution of small objects provides
information on the bulk material properties of the objects,
\citep{Obrien2003,Kenyon2004} while the size distribution of
large objects can provide information on the conditions under which
these bodies formed \citep{Kenyon2002}.

We performed a 3.0 square-degree survey of the Kuiper belt to
determine the size-distribution of large ($D \gtrsim 50$ km) Kuiper
belt objects (KBOs) via a measure of the belt's luminosity
function. This is the deepest photometric Kuiper belt survey of this
size ever completed.

In section 2 we describe our observations. In section 3 we describe our search technique and data
analysis. In section 4, we derive a relation between the
size-distribution and the luminosity function. In section 5, we
describe the statistical analysis used. We present our results 
 in section 6. In section 7 we discuss the implications of our findings, and in section 8 we present our conclusions.

\section{Observations\label{sec:2}}
The observations used in our survey were taken with the 3.6 m
Canada-France-Hawaii Telescope (CFHT) and the
Cerro-Tololo-Interamerican Observatory (CTIO) 4m Blanco
telescope. Observations at CFHT were acquired with both the CFH12k (0.33 square degree fov.) and
MEGAPrime (0.88 square degree fov) mosaic cameras while observations at CTIO were made with the
Mosaic2 camera (0.38 square degree fov), providing 0.67, 0.84, and 1.54 square degrees of searchable area respectively for a total of 3.0 square degrees of searchable area. Details of the observations are provided in
Tables~\ref{tab:Fielddata} and ~\ref{tab:pointings}.

All of these observations were originally acquired for use in searches for
satellites of Uranus and Neptune \citep{Kavelaars2004,Holman2004} and prior to this work, none of these fields had been searched for KBOs.
 The observations were made when Uranus and Neptune (and all KBOs in each field) were at or near
opposition, and covered the projected
Hill-spheres of each planet. The CFH12k and Mosaic2 observations  excluded the area closer than
$3\arcmin$ to the planets, to avoid scattered light in the images.
 These
observations all occurred near the ecliptic and are well suited to
a deep search for Kuiper belt objects (KBOs).

Approximately 10-20 bright, non-saturated stars ($\sim 20$ mag.) per chip were
used as reference stars for the image reductions. The variation in the average reference
star magnitudes follow approximately that expected from the varying
airmass of the observations (see Fig.~\ref{fig:MAG_vs_HA}) indicating photometric conditions during the observations. The remaining scatter is $\sim 0.02-0.04
$ mag, significantly less than the shot noise for the brightest objects
detected.

\section{The Data Processing and Image Search\label{sec:3}}

To determine the behavior of KBO size
distribution requires the discovery of a large number of faint moving
sources.
  If long exposures are used, then the sources will move far
enough (more than the size of the seeing disk) that a trailed image
results and no additional depth will be achieved.
  In our deep
searches we have adopted a strategy of taking exposures short
enough that trailing losses will be negligible.  We then shift the
individual exposures to account for the expected sky motions of the
objects of interest.
  These shifted images are then combined
together to achieve the depth needed.
  To account for the
various possible sky motions of KBOs, we have shifted the images at a
variety of rates and angles and then {\it visually} examined each of the
combined images (stacks), searching for point like sources (see Fig.~\ref{fig:shifts}).

\subsection{Data Preprocessing}

\begin{description}
\item[MEGAPrime:] CFHT provides wide field imaging data from the
  MEGAPrime camera in a `preprocessed' format, ready
  for science exploitation. The frames provided have been processed
  using the CFHT ELIXIR/FLIPS \citep{Magnier2004} processing system. As
  part of this processing, unique bias, flat-field, fringe, and
  scattered light images are produced for each `camera-run' (typically
  matching a ~14 night dark-run) and applied to all frames acquired
  during that camera run. Dark runs are broken into multiple
  camera-runs if a significant change in the camera performance or
  image characteristics is detected. All the MEGAPrime images used in
  this project were acquired within a single camera-run and have
  been `detrended' with a common set of calibrator images.  Standard
  calibration from CFHT results in a flux conserved image with
  constant sky level across the image that has a typical variation of
  $\pm 3\%$.

\item[CFH12K:] These data were originally acquired as part of a search
  for irregular satellites of Uranus \citep{Kavelaars2004} and are different from the data used for the wide field satellite search presented in \citet{Petit2006}.  Due to
  the strong time constraints of the imaging, the images were
  acquired in `classical' observing mode and not automatically
  detrended by the CFHT queued service observing team.  In November 2004 the Canadian
  Astronomical Data Center (CADC) acquired the ELIXIR/FLIPS software
  and calibrators for the CFH12K data and subsequently `detrended' the
  entire set of CFH12K images for which global calibrator frames
  (bias, flats, fringe, scattered light) were available.  Part of the
  re-processing effort included the re-processing of the CFH12K frames
  used in this project.  These reprocessed images were used in this search and provide a sky flatness
  of typically $\pm 4\%$.

\item[CTIO:] The CTIO images were originally acquired as part of a
  project to search for satellites of Neptune \citep{Holman2004}. 
Bias
  frames were acquired on each observing night and a combination of the
  overscan strip and an average of a dozen bias frames was used to remove
  the instrumental ADC bias from each frame. 
 During the period of
  observations, a number ($\sim 15$) of independent fields were
  observed in order to measure the astrometric positions of some
  previously-known Kuiper belt objects. 
 These sequences of fields
  were `median combined' using the IRAF images.median task with high
  pixel clipping. 
 These combined images provided an excellent
  flat-field frame that was divided into the search images using the
  IRAF \citep{Tody1993} mscred.ccdproc task. 
 This process resulted in images
  which have flat sky across the entire mosaic with a typical scatter
  of $\pm 3\%$ in the sky flux level. 
\end{description}

The curvature of the focal plane for all instruments used in this project was small enough compared to the spatial shifts applied to the images during image processing, such that spatially flattening the images was unnecessary.

For the MEGAPrime and CFH12k data, zeropoint calibrations were done with the
standard Elixir routines and were provided by the CFHT Elixir QSO \citep{Magnier2004} and were reported to be $Z_{field,g'} = 26.46 \pm 0.02$ mags. and $Z_{field,R} = 26.22 \pm 0.02$ mags. Hence these magnitudes are presented in the Sloan g' \citep{Smith2002} and Kron-Cousins R. For the N10032W3, N11033, NEP0813NW3, and NEP0815NE3 fields, calibration frames were unavailable, and a nominal zeropoint for the CTIO mosaic of $Z_{VR}=26.0$ was used presenting the magnitudes of the objects discovered in these frames in the VR filter.

\subsection{Artificial Object Planting}

To determine the search efficiency as a function of magnitude, artificial objects representative of KBOs were added (implanted) in the images.

The sky rate of motion ($\dot{\theta}$) of an object on a circular orbit observed at opposition on the ecliptic, at heliocentric distance $r$ can be approximately given by

\begin{equation}
\dot{\theta} \sim 148 \left[\frac{1}{\Delta}-\frac{1}{r^{3/2}}\right]  \mbox{ arcsec. hr$^{-1}$}.
\end{equation}

\noindent
We implanted 100-150 moving sources into each CCD image. All sources were added blindly to the data before searching began; the artificial source lists were revealed to the operator only after searching was completed. The rates and angles of
motion of the artificial sources (approximately $1.3-4.1$ arcsec. hr$^{-1}$ and $\pm 10^o$ from the ecliptic) were typical of objects on
circular orbits at heliocentric distances from $25 - 100$ AU with inclinations between $0-70^o$. 
Each implanted source was given a
randomly selected flux, equivalent to that of a source between $23-27$
mag. The distribution of artificial sources is shown in
Fig.~\ref{fig:r_i}.  Additionally, five artificial sources with flux
levels equivalent to $21$st magnitude sources were implanted, with zero
inclination. These objects have sufficient flux to allow us to flag
errors in the image combining algorithms (failed image subtraction,
wrong mask limits, etc.) in advance of searching the data, and
provided reference moving sources for computing aperture corrections in the
final image combinations.

To account for  image-to-image flux variations due to changes in airmass and possible sky transparency,    the flux of the planted artificial objects was varied with respect to a reference image (usually the middle image in the list) to match the average flux variation of the reference stars (see Fig.~\ref{fig:MAG_vs_HA}).

The point spread function (PSF) for each frame was approximated on a per-chip basis using the stellar profile of a single bright isolated star. The PSF variation across individual images was small enough that the use of a PSF that varied with position was unnecessary. 

Artificial sources whose sky motion would result in a drift of the centroid of the PSF by more than one pixel during an exposure were represented by a series of fainter sources. The number of  faint objects was equal to the number of pixels the source would drift during the exposure. The total flux of all the fainter sources was equal to the flux of the original source. In this way, we fully included trailing effects into the implanted sources, and thus accounted for the effects of trailing in our final search efficiencies.

\subsection{Image Subtraction \label{sec:3.3}}

  During our
previous deep searches \citep{Gladman1998,Gladman2001,Holman2004,Kavelaars2004} we found that the two largest inhibitors
to this search method are trails in the combined images (caused by
bright stars) and the enormous human effort required to visually
examine the broad range of rate/angle combinations needed to ensure detection
of KBOs. To combat these issues, we utilize an image subtraction routine to remove most stationary background sources and implement a new image display method which greatly eases the strain on the user during the visual image search. 
 We describe the image subtraction routine, and the image display method here.

A template image was subtracted from each image to remove stationary
sources from individual images prior to being stacked, thus
improving our ability to detect moving sources. To create the image
subtraction template, an artificial skepticism 
weighted average of all images \citep{Stetson1989}, with the artificial moving sources
already added, was created. This method creates a per-pixel weighted
average, with weights calculated iteratively, and quickly converges on
an average which places very little weight on spurious pixels such as
cosmic ray hits.

For our image subtractions, we chose to use psfmatch3 developed by one of us (CJP) for the Supernova Legacy Survey \citep{Pritchet2005, Astier2006}. This subtraction routine compensates for variations between the image quality of the template image and that of the individual images. Several programs that perform this task already exist (see \citet{Alard2000} and references therein). Psfmatch3 has several advantages. The subtraction kernel can have arbitrary form, and does not require representation by  a set of basis functions to perform the subtraction. In addition, it can automatically remove both spatial and background variations between images. The result of the subtraction routine, is smooth subtracted images with zero average backgrounds. A detailed description of the method is presented in Appendix~\ref{ap:psfmatch3}.

While the image subtraction removed any stationary non-saturated sources, cosmic ray spikes and other spurious hot pixels remained. To compensate for this, an image mask was applied, such that if a pixel had a value outside a chosen range, then a 3 pixel by 3 pixel box about that pixel was set to have zero value. The lower limit to the range was chosen to be -5 times the background noise (the average background was set to zero in the image subtraction). The upper limit was chosen to be $\sim 25 000$ counts, such that most saturated regions and cosmic ray spikes were masked out of the data. This procedure masked the centres of the brightest KBO sources. This however, did not hinder the detections of these sources  as they were still glaringly obvious even in individual frames.

\subsection{Image Stacking\label{sec:3.4}}
The subtracted, masked images containing artificial sources were shifted before they were stacked,  such that each subsequent image was spatially shifted to compensate for the predicted motion of KBOs in the frame.  Sources whose sky motion was well matched by the shifts applied to the images appear nearly round in the stack (see Fig.~\ref{fig:shifts}) while any residual flux from stationary sources was trailed.

If a source's rate of motion was not well matched by the spatial shifts applied to the subtracted images, that source appeared extended in the stack. This characteristic trailing provided a very robust means of discriminating between real sources and noise. Noise sources were produced during the shifting and stacking when positive flux regions from different images were caused to overlap by the choice of spatial shifts. These false sources, however, did not show the image shape variation characteristic of a real source; false sources did exhibit trailing, but of a different length and width compared to a real (or artificial) source. As such, false sources were not selected as candidates by the trained operator.

The template image used for the Psfmatch3 subtraction process contains both the real and artificial moving sources. These appear as faint trails in the template image. Subtracting this template from the input images results in a low flux area behind each moving source. This feature only occurs around the positions of real (and artificial) moving sources and provides an additional and robust means of source-noise discrimination (see Fig.~\ref{fig:shifts}) and was required to be present in order to mark a source as a candidate.

The quality of the image subtraction, and hence the final searchable images suffered from a few effects: bad columns in the CCD, bleeding from saturated stars, and regions of poor subtraction around bright galaxies. These problems were exacerbated by image shifting; the unsearchable area of these regions were expanded, reducing the overall searchable area. Gradients of the background caused by bad image subtraction around bright sources, produced images that were difficult to display, reducing the search efficiency. Our efficiency of detecting artificial, planted sources accounts for all reductions in area as the artificial sources were planted at random locations occupying the full spatial extent of each CCD image.

\subsection{Visual Search}
To search the combined stack of shifted images, the stacks were divided up into $\sim 200$ image subsections with a 20 pixel overlap between neighboring subsections. One-by-one, a grid of rates and angles like that shown in Fig. \ref{fig:shifts} was displayed for each image subsection. Each grid was searched by eye, and sources were recognized by their characteristic trailing and subtraction wells. We found that a five-by-five grid of shift rates and angles maximized the detection efficiency while minimizing the time spent searching. The low variations in the sky background, resulting from the template subtraction, and an image display tool developed specifically for this searching allowed us to rapidly search many data sets.

The pixel coordinates of potential candidates were recorded along with the rate and angle combination that produced the most circular image. A candidate was selected as a planted object if its  marked image location was within 10 pixels of an artificial source location. The list of detected artificial candidates was used to determine the detection efficiency as a function of artificial source brightness. The detection efficiencies for each chip of a given field were averaged together to provide a global, per field, detection efficiency which we then modeled (see~Fig.~\ref{fig:eff} and Table~\ref{tab:effs}) using the equation

\begin{equation}
\eta\left(m|\eta_{max},m_\star,g\right) = \frac{\eta_{max}}{2}\left[1-\tanh\left(\frac{m-m_\star}{g}\right)\right].
\label{eq:eff}
\end{equation}

\noindent
Our deepest and widest field has a $\eta(m)=50\%$ threshold at $m(g')=26.4$ and a sky coverage of $0.84$ square degrees. The sky coverage of our combined data is 3.0 square-degrees. 

Remaining candidates not marked as artificial source detections were re-examined at a finer grid of 8 rates  and 8 angles to further discriminate between real sources and noise. Candidates were rejected as false positives during this process if the variation of their image shape and trailing length with rate and angle did not match that of brighter sources with similar apparent rates of motion, whose detection was more robust. While this approach allowed us to isolate and remove false detections, the artificial sources were treated differently from the real sources, as we did not examine candidates marked as artificial sources in the finer rate and angle grid.  The brightnesses of all false positives were found to be faintward of the 50\% threshold of their discovery field, when their fluxes were measured in the same way as all real detections. Therefore, below the 50\% threshold, our search efficiency was no longer representative of the true search efficiency, and subsequently, sources fainter than the 50\% threshold of each field are ignored in our analysis.  [ NOTE: Here we define the 50\% threshold as the point where $\eta\left(m\right) = 50$\%.]

A series of follow-up images of the MEGAPrime field as well as the NEP0813NW3 CTIO field were obtained one or two nights after the initial discovery images.  Using the short first night discovery arcs, we projected the motion of each detection forward to the time of the follow-up observations in order to predict the sky location of the source on the second night's images. For 3 of the 25 sources in the MEGAPrime field, the predicted location placed the sources in the gaps between the CCDs of the MEGAPrime MOSAIC.  Aside from these 3 sources, all of our initial detections brighter than the 50\% threshold were confirmed. Only the faintest initial detection in the MEGAPrime field was not confirmed on the second night, likely due to poorer average seeing conditions during the second nights observations. All 6 detections in the NEP0813NW3 field were successfully confirmed on the second night.

None of our other search fields had follow-up observations.  We were confident, however, that all detections in the other fields brightward of the 50\% threshold were real because all detections brightward of the 50\% threshold (excluding those that fell on chip gaps) in the MEGAPrime and NEP0813NW3 fields were confirmed, and all fields were processed and searched with identical procedures.

For the range of rates and angles that we planted objects into the data, we found no significant variation in detection efficiency versus rate or angle of motion for any of our fields (see Fig. \ref{fig:r_i}).  This is important to note because any significant variation in efficiency with rate of motion needs to be included when deriving the size distribution from the observed luminosity function.  We did not search for sources with rates of motion consistent with objects at distances further than 100 AU as the search method we employed is not sensitive to sources beyond this distance; such distant objects would not show enough apparent motion over the time-span of our observations to detect the faintest sources, and a degradation of the detection efficiency with distance would occur. Therefore, our determination of the luminosity function only applies to  KBOs closer than 100 AU.

\subsection{Source flux measurements and detection confidence \label{sec:flux_meas}}

To measure the flux from each detected source, we stacked the non-subtracted images containing the artificial objects. These images were shifted at the rate and angle that produced the roundest image for the source whose brightness was to be measured. Three sets of stacks were produced, by averaging the images, while using a pixel by pixel cut that threw away pixel values outside 3-sigma of the mean value. Three averages were made from images from the first half, middle half, and last half of the observing period for each field.  These average images provided three separate, mostly uncorrelated, measurements of each source's flux. Aperture photometry was performed using the IRAF daophot.phot task, set to use an aperture of 4 pixels radius (close to the size of the FWHM of point-sources in our images). Using the five 21st magnitude planted `reference' objects, and the mkapfile IRAF routine, aperture corrections were determined for each of the three image combinations (first, middle last) and  were used to correct individual flux measurements for each source. The IRAF mkapfile task reported our aperture corrections to an accuracy of $\sigma_{aper} \lesssim 0.03$.  The magnitudes, as measured on each exposure set, were averaged for the final reported magnitudes (see Table~\ref{tab:detections}).  We did not measure the flux in a particular stack for sources that were within a few seeing disks of bright or bloomed stars or galaxies ($\sim$ 1 in 10 possible source measurements), in that particular stack. 

For two objects in the MEGAPrime data, the measured magnitudes varied significantly more than the uncertainty of each individual measurement, indicating a possible light curve.  For these two objects, the magnitudes measured from the first image stack were used in our determination of the luminosity function, as was done in \citet{Bernstein2004}. The second night's data provided an additional 3 magnitude measurements of any followed-up source.  For all but the variable sources, we used the average magnitude measured from all image stacks including those from the second night. 

\subsection{Characterization}

As required by the maximum likelihood routine, we parametrize the magnitude uncertainty and detection efficiency of our survey. The scatter between each artificial source's measured and inserted magnitudes was used to determine the precision of our flux measurements.

For background limited sources, the uncertainty of a source's measured magnitude can be represented by 

\begin{equation}
\Delta m = \gamma 10^{\left(m-Z\right)/2.5} \label{eq:dm_final}
\end{equation}

\noindent
where $Z$ is the telescope zeropoint. $\gamma$ is a function of the background $b$, the camera gain $g$, and the on-source integration time, and is fit to the scatter of the artificial source's measured and inserted magnitudes. The fit values of $\gamma$ for each field can be seen in Table~\ref{tab:effs}. The magnitude measurement scatter for each source was then drawn from Eq.~\ref{eq:dm_final} using these best fit values.

The theoretical value of $\gamma$ is 

\begin{equation}
\gamma^*=\frac{2.5 }{\ln 10}  \sqrt{\frac{\pi r^2 bg}{tN}}
\label{eq:gamma_final}
\end{equation}

\noindent
where $N$ is the number of exposures. If we use the typical values from our MEGAPrime observations, we expect a value of $\gamma^* = 0.12$. The measured value of $\gamma=0.27$ is clearly larger. \citet{Newberry1991} however, has shown that the noise calculation from Eq.~\ref{eq:gamma_final} is incomplete, and does not fully account for the noise introduced into observations during the data reductions. They found that the noise estimate given in Eq.~\ref{eq:gamma_final} can be too small by a factor of 2. Similarly, source fluxes were measured off of images that were shifted and stacked. If the shifts were slightly different than the source's true motion, the measured magnitude would differ slightly from the true magnitude. Thus we find that the uncertainties measured from the artificial sources are close to the expected values. The net uncertainties used in our luminosity function determinations are the 1-sigma shot-noise, $\Delta m$, aperture correction $(\sigma_{aper} \lesssim 0.03)$, and calibration uncertainties, added in quadrature.

\section{The Luminosity Function}

The size distribution of the KBOs can be determined directly from their luminosity function. 
The relation between the luminosity function and the size distribution has been derived previously (see, for example, \citet{Gladman2001}). We reproduce the relation and discuss complications due to possible variations of KBO albedos, using a slightly different approach than that in \citet{Gladman2001}. 

Consider the number of objects $N$ in a survey field, given by

\begin{equation}
N=A\int\int R\left(r\right) S\left( D\right) dD dr
\label{eq:Nbasic}
\end{equation} 

\noindent
where $R\left(r\right)$ and $S\left(D\right)$ are the radial and size distribution functions, $r$ and $D$ are  object heliocentric distance and diameter, and $A$ is some normalization constant. We model the size distribution as $S\left(D\right)\propto D^{-q}$.  
The magnitude $m$ of an object at heliocentric distance $r$, geocentric distance $\Delta$, with diameter $D$ and albedo $p_\lambda$ is 

\begin{equation}
m=K_\lambda + 2.5 \log_{10}\left( r^2 \Delta^2 D^{-2} p_\lambda^{-1} \right),
\label{eq:magnitude}
\end{equation}

\noindent
where $K_\lambda$ and $p_\lambda$ are wavelength dependent; in R-band, $p_R\sim4\%$ gives $K_R\sim 18.8$.  The smallest observable object at a given distance $r$ in a survey with limiting magnitude $m_{max}$, has diameter $D_{min}=p_{\lambda}^{-1/2} r^2 10^{(K_\lambda-m_{max})/5}$ where we have made the approximation $r\equiv \Delta$. Similarly, the largest observed object at distance $r$ will have diameter $D_{max}=p_{\lambda}^{-1/2} r^2 10^{(K_\lambda-m_{min})/5}$.

Using these limits and integrating, Eq.~\ref{eq:Nbasic} becomes

\begin{equation}
N(m<m_{max}) =  A \int \frac{r^{2(1-q)}}{p_{\lambda}^{(1-q)/2}} R(r) dr \mbox{ } \frac{\left[10^{(1-q)(K_{\lambda}-m_{min})/5} -10^{(1-q)(K_{\lambda}-m_{max})/5}\right]}{1-q}
\label{eq:Nfinal}
\end{equation}

\noindent
if $q \neq 1$. Note: this assumes that $D_{min}$/$D_{max}$ are not the smallest/largest objects in the belt; if this were the case, using $D_{min}$/$D_{max}$ as the integration limits as we have defined them here is incorrect.

For $q>1$ and a survey where $(q-1)\left(\frac{m_{max}-m_{min}}{5}\right) \geq 2$, we have $10^{(1-q)(K-m_{min})/5}  \ll 10^{(1-q)(K-m_{max})/5}$, and the cumulative luminosity function is given by

\begin{equation}
\begin{array}{cl} 
N(m<m_{max})  & \simeq A \int \frac{r^{2(1-q)}}{p^{(1-q)/2}} R(r) dr \mbox{ } \frac{10^{(1-q)(K-m_{max})/5}}{1-q} \\
       & \simeq \frac{A}{(1-q)} \int \frac{r^{2(1-q)}}{p^{(1-q)/2}} R(r) dr \mbox{ } 10^{\alpha(m_{max}-K)} \\
\end{array}
\label{eq:Nfinalt2}
\end{equation}

\noindent
where we have substituted $q=5\alpha+1$.

The observed cumulative luminosity function is well represented by a power-law of the form

\begin{equation} 
N(m)=10^{\alpha (m-m_o)}.
\label{eq:N}
\end{equation}

\noindent
which gives the number of KBOs per square-degree,  on the belt mid-plane, brighter than magnitude $m$. Comparing this with Eq.~\ref{eq:Nfinalt2}, reveals that the luminosity function slope $\alpha$ and the size distribution slope $q$ are related by

\begin{equation}
q=5\alpha+1
%\alpha=\frac{q-1}{5}.
\label{eq:alpha-q}
\end{equation}

From Eqs.~\ref{eq:Nfinalt2} and~\ref{eq:alpha-q} we see that, for a size distribution that is independent of distance, the choice in radial distribution is not significant and only affects the interpretation of the normalization of the observed luminosity function, not the inferred size distribution slope.

Here we have assumed that the KBO albedos do not vary with distance, or size. If the distribution of KBO albedos varies only with distance, then this will only affect the normalization of the luminosity function.

The albedo of Pluto and Eris ($D \gtrsim 2000 km$) are $p_{Pluto}\sim 0.6$ and $p_{Eris} \sim 0.6-0.86$ \citep{Young2001,Bertoldi2006,Brown2006}  while smaller objects ($D \sim 100$ km) have
been shown to have albedos $p\sim 0.06$ \citep{Grundy2005}. These data suggest an increase in albedo with size. We can understand the effects of such a trend on the interpretation of the LF by considering  a toy model where KBO albedos vary as $p \sim D^{-\beta}$.  This functional form retains the analyticity of Eq.~\ref{eq:Nbasic}, and reveals the effects of albedo variations on the inferred size distribution. Under this assumption, Eq.~\ref{eq:Nfinalt2} becomes

\begin{equation}
N(m<m_{max}) = A \int \frac{r^\frac{4(1-q)}{2-\beta}}{1-q} R(r) dr 10^{\frac{(K-m_{max})(1-q)}{2.5(2-\beta)}}.
\label{eq:Nnasty}
\end{equation}

\noindent
Thus the slope of the size distribution is

\begin{equation}
q=5 \alpha \left(1-\frac{\beta}{2}\right)+1
\label{eq:alpha-qnasty}
\end{equation}

\noindent
We see that in the case of constant albedo $(\beta=0)$ we get back Eq.~\ref{eq:alpha-q}.

Current albedo data imply $\beta \sim -1$ \citep{Stansberry2007}.  Thus, an estimate of $q$ which assumes $\beta=0$
potentially under-estimates the steepness of the intrinsic size distribution. Our knowledge of the relation between albedo and size however, is currently insufficient to constrain $\beta$.

The reader is cautioned that the current determinations of the size distribution from the Kuiper belt luminosity function are based on a few poorly constrained estimates of KBO physical properties and the inferred slope $q$ is probably underestimated.

\section{Maximum Likelihood fits \label{sec:MLF}}

To determine the optimal values of $\alpha$ and $m_o$, we use a
maximum likelihood method. We extend the maximum likelihood analysis of \citet{Gladman2001} to account for
observations made in different wave-bands, as well as systematic
differences in the normalization, $m_o$, that may affect the results of
combining separate data sets, such as systematic calibration errors, sky density variations, and variations in average colour of the observed KBOs compared to the average colour of all KBOs. 

Given that the probability of detecting an individual KBO is independent of detecting the next, the likelihood function for a single
survey $k$ takes the form

\begin{equation}
L_k \left(\alpha, m_o | m_1, m_2, ...\right) \propto \exp^{ - \tilde{N_k}} \prod_i P_i
\label{eq:L_basic}
\end{equation}

\noindent
where $m_i$ is the magnitude of detection $i$, $\tilde{N_k}$ is the number of objects expected to be detected in a given survey, and $P_i$ is the probability of having object $i$ given the underlying luminosity function. $\tilde{N_k}$ is given by

\begin{equation}
\tilde{N_k} = \int dm \  \Omega \ \eta(m) \ \Sigma(m|\alpha, m_o)
\label{eq:Nk}
\end{equation}

\noindent
where $\Omega$ is the survey area, $\eta(m)=\eta(m|\eta_{max},m_*,g)$ is the detection  efficiency for an object with magnitude $m$, and $\Sigma(m|\alpha,m_o)$ is the differential density of objects on the sky. $P_i$ and $\Sigma(m|\alpha,m_o)$ are given by

\begin{equation}
 P_i=\int dm  \ \Sigma(m|\alpha,m_o) \ \epsilon_i(m).
 \label{eq:Pi}
\end{equation}

\noindent
and

\begin{equation}
 \Sigma(m|\alpha,m_o) =\frac{dN(m)}{dm} = \ln(10) \ \alpha \ 10^{\alpha (m-m_o)},
 \label{eq:diff_lum_func}
 \end{equation}
 
 \noindent
where $\alpha$ is the slope of the luminosity function (LF), $m_o$ is the magnitude at which the sky density is 1 KBO per square degree, and $\epsilon_i(m)$ is a functional representation of the photometric uncertainty for object $i$. A formal derivation of this likelihood function is presented in \citet{Loredo2004}. 

We choose to represent the magnitude uncertainties as gaussian. ie.  
 
 \begin{equation}
 \epsilon_i(m)=\frac{1}{\sqrt{2\pi \Delta m_i^2}}e^{\frac{-m^2}{2\Delta m_i^2}}
 \label{eq:error_m}
 \end{equation}
 \noindent
 where $\Delta m_i$ is the uncertainty in the magnitude measurement of each object. This treatment of uncertainties for faint sources is incorrect, but is a sufficient approximation for detections brightward of the 50\% efficiency threshold. Brightward of this threshold, the gaussian approximation will not affect the results of the maximum likelihood inference \citep{Bernstein2004}.

For a group of surveys with calibration uncertainties and colour offset variations much smaller than the uncertainty in the flux measurements of the brightest objects detected, the net likelihood resulting from combining multiple surveys together is the product of the likelihoods of each individual survey. In reality, photometric calibration and colour offset uncertainties are not insignificant, and therefore, need to be considered when combining different surveys.  

Additionally, the apparent sky density of KBOs and, hence, $m_o$ are strong functions of latitude and longitude (Kavelaars {\it et al.}, In Press). When combining separate surveys, differences in the flux-limited sky densities will skew the inferred slope of the LF. 

We account for these effects with the addition of two new parameters for each survey, the colour parameter $C_k$, and the density parameter $\Delta m_{o,k}$. Eqs.~\ref{eq:Nk}, \ref{eq:Pi}, and \ref{eq:diff_lum_func} become

\begin{equation}
\tilde{N_k} = \int dm \  \Omega \ \eta(m|\eta_{max},m_*,g) \ \Sigma_k(m-C_k|\alpha, m_o, \Delta m_{o,k}),
\label{eq:NkC}
\end{equation}

\begin{equation}
 P_i=\int dm  \ \Sigma_k(m-C_k|\alpha,m_o,\Delta m_{o,k}) \ \epsilon_i(m),
 \label{eq:PiC}
\end{equation}

\noindent
and 

\begin{equation}
 \Sigma(m-C_k|\alpha,m_o,\Delta m_{o,k}) = \ln(10) \ \alpha \ 10^{\alpha (m-C_k-(m_o-\Delta m_{o,k}))}.
 \label{eq:diff_lum_func_C}
 \end{equation}
 
 \noindent
As can be seen, for the case of a power-law LF, the parameters $C_k$ and $\Delta m_{o,k}$ are degenerate. For a non power-law model however, $C_k$ and $\Delta m_{o,k}$ are no longer degenerate, and need to be treated separately. As we consider non power-law LFs in this manuscript, we choose to treat $C_k$ and $\Delta m_{o,k}$ as different parameters when evaluating the likelihood.

This treatment substantially increases the number of parameters in the fit, while not providing any new information about the shape of the LF. Thus we treat these additional parameters as nuisance parameters, and marginalize (integrate) the likelihood equation over the expected ranges of each parameter. The final likelihood equation when combining $\mathcal{N}$ surveys is

\begin{equation}
L( \alpha, m_{o},C_1,C_2,...,C_\mathcal{N},\Delta m_{o,1},\Delta m_{o,2}, ..., \Delta m_{o,\mathcal{N}} ) \\
= \prod_{k=1}^{\mathcal{N}} \int \int L_k(\alpha, m_{o}, C_k, \Delta m_{o,k}) dC_k d\Delta m_{o,k}.
\label{eq:LF_final}
\end{equation}

\section{Results \label{sec:Results}}

Each detection in our survey is listed in Table~\ref{tab:detections} along with estimated barycentric distance and inclination determined from fit\_radec \citep{Bernstein2000}. The fit\_radec routine determines a set of orbital parameters that best fit the observations in a least-squares sense. Because of the large degeneracies between orbital parameters for short-arc observations such as those presented here, fit\_radec initially assumes that the objects are near perihelion, and are on bound, nearly circular orbits, near the ecliptic plane. The routine  determines 2-sigma uncertainties, which contain 95\% of the orbits which are consistent with the measured positions of the objects. In this survey, we have detected 72 KBOs, 53 of which are brightward of the 50\% threshold of our search, and these 53 detections are used in our maximum likelihood analysis.

As the UNE and UNW fields are close on the sky,  observed together through the same filter/telescope, and the skies were photometric during those observations, we combine the two fields for the maximum likelihood fits. Similarly, we treat the N10033 and N10032W3 and the NEP0815NW3 and NEP0815NE3 fields in the same fashion. We refer to these combined fields as the UN, CTIO01, and CTIO02 fields. The CTIO01 and CTIO02 fields were not combined together, as they were observed in separate years.

To extend our results, we include the observations from other surveys that have well-measured efficiency functions for each field in the survey. We define the F07 sample as those objects detected in the surveys from \citet{Jewitt1998}, \citet{Gladman1998}, \citet{Chiang1999}, \citet{Gladman2001}, \citet{Allen2002}, and \citet{Petit2006}, as well as those detected in the UN, CTIO01, and CTIO02 fields. 

In the F07 sample we also include all on ecliptic detections in \citet{Trujillo2001b} (T01) in fields with detection efficiencies classified as `good' or `medium'. Because the T01 images cover a large range of latitudes, longitudes, and detection efficiencies, we divided the survey into smaller ``sub-fields'' to avoid large density variations from field to field. Each sub-field spans $\sim$1 - 1.5 hrs. RA. We do not include any of the high-latitude data from T01 as the KBO latitude distribution is not sufficiently understood to be able to predict the density of KBOs at high latitudes and we are thus unable to provide a reasonable estimate of $\Delta m_{o,k}$ for the high-latitude fields. We avoid complications due to detection efficiency variations by further sub-dividing the data from T01 into separate groups based on the good and medium detection efficiencies defined in that work. We do not consider any T01 fields with detection efficiencies classified as `poor'. The sub-field divisions listed by field ID (see Table 2 in Trujillo et al. 2001) are presented in Appendix B.  

We include all fields from the above surveys, even those with few or no detections in the F07 sample. This is the sample on which all our LF analysis was performed. 

We chose to exclude the data from \citet{Bernstein2004} as the intent of this work is to determine the shape of the bright end of the luminosity function.

When computing our LF estimate, we ignored all detections fainter than the $\eta(m)=50\%$ detection threshold of each individual survey, and truncated (``cut") the detection efficiency to zero faintward of that threshold. Correctly determining the efficiency function at low levels is notoriously difficult and prone to error. Effects such as Malmquist-bias further complicate detection efficiency measurements, at low brightnesses. 

To test for a bias in the fits caused by an efficiency truncation, we ran a series of Monte-Carlo realizations that simulated the observation of a luminosity function from a single survey. We also performed this analysis using a set of surveys with parameters like those in the F07 sample. We performed 2000 random realizations for efficiency cuts ranging from 0-90\%. We found that, for a single survey, the mean of the estimated LF slope is biased to steeper slopes when truncating the detection efficiencies at some non-zero threshold.  The bias was found to decrease asymptotically to zero as the efficiency threshold was moved faintward. 

We found the bias caused by including an efficiency cut is removed however, when combining multiple surveys if the separate surveys have different detection efficiencies. Most detections in a survey occur where $\eta(m)\sim80\%$.  Thus, when two or more surveys are well separated in their limiting magnitudes, so are their detections, which creates a ``lever arm" that dominates the slope determination, effectively  removing any efficiency cut bias in the fit. 

For each field with 5 or more detections which is the minimum number of detections required to provide a reliable measure of the LF, we determined the best-fit parameters for the LF using Eq.~\ref{eq:L_basic}  (see Table~\ref{tab:results}). The 1, 2, and 3-sigma credible regions of these fits are presented in Fig.~\ref{fig:likelihood}. The 1-sigma uncertainties of the best-fit parameters, taken as the extrema of the 1-sigma credible regions, are presented in Table~\ref{tab:results}. Presenting the fit parameter uncertainties in this way describes the full range of allowed values for each parameter individually. These uncertainties are somewhat misleading as they do not describe the correlation between $\alpha$ and $m_o$ (see Fig.~\ref{fig:likelihood}).

The average (V-R) colour of KBOs is $<V-R> = 0.56$ \citep{Hainaut2002}.   Using the VR and R magnitude transformation given by \citet{Allen2001}, as well as the relations between V, R, r' and g' given by \citet{Smith2002}, we determined the expected offsets between the various filters used in the surveys considered here. We found that the average colours of KBOs are $<VR-R>=0.03$, $<r'-R>=0.26$, and $<g'-R>=0.95$.  Using these colours to offset the separate fields to R-band, we find the estimates of $m_o$ provided independently by each survey are not consistent with a single value (see Fig.~\ref{fig:likelihood}). Small shifts in $m_o$ are necessary to make all fields agree at the 1 - 2 sigma level. This is acceptable when considering possible sky density variations, and photometric calibration errors (see Section \ref{sec:MLF}) and justifies the more complicated likelihood equation given by Eq.~\ref{eq:LF_final}. We therefore shifted all data to R-band using the expected average KBO colours, and used Eq.~\ref{eq:LF_final} when determining the best-fit LF parameters.

To determine the possible range in $m_o$ values between surveys, we consider a toy model of the Kuiper belt. As the majority of the objects observed in the surveys considered in this work are Plutinos and classical belt objects (CKBOs), we only consider these objects in the model. We represent the LFs of both populations as power-laws with the same slope, but with different observed sky densities. Then the observed cumulative LF (assuming constant detection efficiencies)  is a power-law, and is given by

\begin{equation}
N_{obs}=10^{\alpha(m-m_{o,c})}+f10^{\alpha(m-m_{o,p})}\equiv 10^{\alpha(m-m_o)}
\label{eq:N_plutinos}
\end{equation}

\noindent
where $m_{o,c}$, $m_{o,p}$, and $m_o$ are the magnitudes at which one object per square degree is observed for the CKBOs, the Plutinos, and all populations respectively and $f$ is the ratio of the number Plutinos to CKBOs.

The mean eccentricity of Plutinos is $\bar{e} = 0.15$, and they are $\sim30\%$ as numerous as the classical belt objects of the intrinsic population. ie. $f=0.3$ (Kavelaars et al. (In press)). The magnitude of an object as a function of heliocentric distance $r$ is given by $m\sim K+10.0 \log(r)$. Thus the variation in $m_{o,p}$ for Plutinos at perihelion versus aphelion is $\sim 1$ mag.
If all CKBOs are observed at $r=43$ AU, then the difference in $m_o$ between fields that observe Plutinos at perihelion versus aphelion is $\sim 0.8$ mags. This is an upper estimate of the true offset because not all Plutinos come to perihelion at once, but provides a reasonable range of integration for each $m_{o,k}$ parameter.

Latitude differences between surveys will cause variations in $m_o$ requiring knowledge of the latitude distribution of the Kuiper belt. While the location of the mid-plane is unclear, it is secure that the mid-plane is inconsistent with the ecliptic \citep{Brown2004,Elliot2005}; there is disagreement about whether the mid-plane is the invariable plane.  \citet{Brown2001} has shown that the KBO latitude distribution above the plane has a Full-Width-Half-Maximum of $\sim 7^o$. Most of the observations considered in the F07 sample are made near the ecliptic. If the true plane of the belt is similar to the invariable plane, or the Kuiper plane proposed by \citet{Brown2004}, then the density variation due to latitude between ecliptic fields is at most 5\% and would affect the $m_{o,k}$ values by $\sim 0.05$ mags. This variation is small compared to that expected from changes in longitude, and is ignored in our analysis. We bound the integration of Eq.~\ref{eq:LF_final} over the  $m_{o,k}$ parameters between $\pm 0.4$ mags.

The standard deviation of the $<V-R>$ colour from the MBOSS data set is $\sigma(V-R) \sim 0.15$ \citep{Hainaut2002}. Hence, the expected variation in average object colour between separate surveys due solely to KBO colour variation is $\sim 0.15/\sqrt{N}$ where $N$ is the number of objects in a survey. For the fields considered in the F07 sample, $N\sim1-15$. Zeropoint calibration errors between separate surveys have been as high as $0.1$ magnitudes. Thus a reasonable range of offsets due to calibrations and colour variation is $\sim 0.2$ mags. This offset can occur in either a positive or negative sense on the measured magnitude. Thus we bound the integration of Eq.~\ref{eq:LF_final} over the $C_k$ parameters between $\pm 0.2$ mags.

Maximizing Eq.~\ref{eq:LF_final} using the F07 sample, we find a best fit slope of $\atotal$ and normalization $\mtotal$. The likelihood contours of the fit are shown in Fig.~\ref{fig:likelihood_all}. 

The factor $e^{-\tilde{N}}$ in Eq.~\ref{eq:LF_final} weights the fit towards the lowest possible number of detections. The $m_o$ that maximizes the likelihood equation given by Eq.~\ref{eq:LF_final} is the maximal value within the range allowed by the $\Delta m_{o,k}$ that best describes each individual field considered.  Thus, the best-fit $m_o$ is not applicable to any one field, but rather is a value typical of all data considered in the fit.

 To test if the fit is consistent with the observations, we employed a series of Monte-Carlo simulations. 
 The simulations involved random realizations of a number of objects drawn from the best-fit power-law model ($\alpha=0.65$, $m_o=23.42$) equal to the number of objects in the F07 sample with the random magnitudes scattered according to our uncertainty model (see Eqs.~\ref{eq:dm_final} and \ref{eq:error_m}). A best-fit power-law of each realization was determined using our maximum likelihood technique, and the distribution of maximum likelihoods was determined from 1000 realizations. We found that the probability of getting a maximum likelihood less than or equal to the maximum likelihood computed from the F07 sample was 43.2\%: the model is fully consistent with these data.

In Fig.~\ref{fig:cumul_diff} ({\it top}) we present a histogram of the differential LF of the F07 sample corrected for detection efficiencies. These data are well described by our best fit. The reader is cautioned however, from drawing conclusions about the LF from this diagram alone. All source magnitudes have been adjusted to R-band using typical KBO colours (see above). The observational data from different surveys however, contain calibration and colour offsets, and the observations have not been adjusted to reflect variations in sky densities, as a standard model of the density variations is not known. The minor discrepancies between the observed LF and the model apparent in Fig.~\ref{fig:cumul_diff} maybe caused by these effects.

In Fig.~\ref{fig:cumul_diff} ({\it middle}) we show the net effective area for all fields as well as a differential histogram of the object magnitudes  shifted to R-band with completeness corrections. The different fields which are included in the F07 sample all have different 50\% thresholds, resulting in  the broad fall off in effective area between $m(R)\sim 21$ to $26$.

In Fig.~\ref{fig:cumul_diff} ({\it bottom}) we show the logarithm of the ratio of the observed and fit differential LFs, as a function of magnitude.  This figure shows that the LF from $m(R)=21 $ to $26$ is well described by a single power-law.

\subsection{Broken Power-law Models}
\citet{Bernstein2004} conclude that the slope of the luminosity function rolls over to shallower values for fainter magnitudes and that the luminosity function is well described by a rolling power-law given by the functional form

\begin{equation}
\Sigma(m)=\Sigma_{23} 10^{\alpha(m-23)+\alpha'(m-23)^2}
 \label{eq:roll}
 \end{equation}
 
\noindent
where $\Sigma(m)$ is the differential surface density of objects, $\Sigma_{23}$   is the surface density of objects at $m(R)=23$, $\alpha$ is the bright end slope, and $\alpha'$ is the slope derivative.  They found that a rolling power-law was a better fit to their observations than the single power-law used here. This suggests that a deviation in the form of a flattening of the power-law might be visible in our data. 

To look for evidence of a roll-over in the KBO LF we fit Eq.~\ref{eq:roll} to the F07 sample. As the density parameter for Eq.~\ref{eq:roll}, $\Sigma_{23}$, is different from $m_o$ in Eq.~\ref{eq:N}, we introduce a density offset parameter $\Delta \Sigma_k$ as a multiplicative factor in front of Eq.~\ref{eq:roll}. We maintain the same range in colour and density offsets as used in the power-law fit (see above) and marginalize $C_k$ and $\Delta \Sigma_k$ over the range $\pm 0.2$ and $\pm 0.25$ respectively. Our maximum likelihood method gives a best-fit  $A=0.79\pm 0.14$, $\alpha=0.74\pm 0.09$, and $\alpha'=-0.03\pm 0.04$. Bernstein determined a best-fit of $A=1.07$, $\alpha=0.66\pm0.03$ $\alpha'=-0.05\pm0.015$ (1-sigma uncertainties were extracted from Fig. 4 of \citet{Bernstein2004}), consistent with our results. The increase in uncertainty of $\alpha$ and $\alpha'$ in our fit versus that from \citet{Bernstein2004} is caused by the marginalization in Eq.~\ref{eq:LF_final}. We feel that our approach provides a more realistic estimate of the true uncertainties. 
 
Additionally we fit a simple broken power-law model by Eq.~\ref{eq:LF_final}. The model LF has a sudden change from the bright-side slope $\alpha_1$ to the faint-side slope $\alpha_2$ at a break magnitude $m_B$, and is given by

\begin{equation}
\Sigma =
\begin{cases}
10^{\alpha_1(m-m_o)} & \text{ if $m<m_B$,} \\
10^{\alpha_2m +(\alpha_1-\alpha_2)m_B-\alpha_1m_o} & \text{ if $m>m_B$.}
\end{cases}
\label{eq:my_Broken}
\end{equation}

 The best-fit parameters from maximizing the likelihood for the broken power-law LF while marginalizing over nuisance parameters $C_k$ and $\Delta m_{o,k}$ as per Eq.~\ref{eq:LF_final} are $m_o=23.2 \pm 0.5$, $\alpha_1=0.69 \pm 0.08$, $\alpha_2=0.57 \pm 0.2$, and $m_B=24.4 \pm 0.7$. 
 
 The improvement of the maximum likelihood value when considering the rolling and broken power-law LFs given by Eqs.~\ref{eq:roll} and \ref{eq:my_Broken} fits is only a few percent over that of the single slope model. Hence the additional degrees of freedom included in the more complicated LFs compared to a single slope power-law does not substantially improve the fit and the higher order models are not statistically warranted in the range of magnitudes considered in the F07 sample. Indeed the best fits of both equations are consistent at the 1-sigma level with no break at all ($\alpha'=0$ and $\alpha_2 = \alpha_1$).

To test at what magnitude a roll over to shallower slopes would be inconsistent with the F07 sample, we made use of Monte-Carlo simulations and the Kolmogorov-Smirnov (KS) test. In these simulations, we simulated the observation of a broken power-law LF of the form of Eq.~\ref{eq:my_Broken}, with $\alpha_1=0.65$ breaking to some faint-end slope $\alpha_2$ at break magnitude $m_B$. The simulated surveys and total number of detections were chosen to match the surveys and number of detections in the F07 sample. We included the effects of variable $m_o$ and colour offsets between surveys by randomly selecting these offsets for each individual survey. 

For a given break magnitude and faint-end slope, a parent population of $\sim 10,000$ objects was generated. This parent population was simulated with a set of random $C_k$ and $\Delta m_{o,k}$ offsets linearly sampled from the marginalization range described above ($\pm 0.2$ for $C_k$ and $\pm 0.4$ for $\Delta m_{o,k}$). Calibration and colour offsets occur in the observations in the F07 sample. The random sampling we implement provides a simple means of generating offsets consistent with the way in which those observations were made.

From the parent population, a sub-sample of objects was bootstrapped and the KS-statistic of this sub-sample compared to the parent population was calculated. This procedure was repeated for 1000 random sub-samples of the parent population. In this way, the KS statistic distribution was bootstrapped from the parent population.

The KS statistic of the F07 sample when compared to the parent population was compared to the distribution of KS-statistics generated from the  bootstrapped samples. The roll-over model ($\alpha_2,m_B$) was rejected if 99.7\% of the bootstrapped KS statistics were smaller than the KS statistic of the F07 sample. This was repeated 25 times using different random realizations of the $C_k$ and $\Delta m_{o,k}$ offsets for each choice of faint-end slope and break magnitude. The repetition was necessitated by the randomness included with variable $C_k$ and $\Delta m_{o,k}$ values; in certain circumstances, a particular set of colour and $m_o$ offsets will enhance the chances of observing a break. A model could not be rejected if the break was not rejected in 2 or more of the 25 simulation repetitions for that model.

Presented in Fig.~\ref{fig:KS} is the 95\% probability contour that a particular break model is rejected by the F07 sample. The contour has been smoothed to remove the effects of course sampling in $\alpha_2$-$m_B$ space. As expected, brightward of a particular magnitude most break models would likely have been detected in the observations. From these simulations, we conclude that the observations cannot reject the possibility of a break in the Kuiper belt LF with break magnitudes fainter than $m_B(R)\sim 24.3$.

\citet{Bernstein2004} find that the LF is well described by the harmonic mean of a steep power-law for bright objects and a shallow power-law for faint objects with both power-laws contributing equally at magnitude $R_{eq}$. They find $R_{eq}\sim 22.8 - 23.6$. This is not however inconsistent with the results of our KS test as $m_B$ and $R_{eq}$ are not equivalent parameters between the two LF models. The reader is cautioned about drawing conclusions from comparison of these two parameters.

While no break is apparent in the observations, the best-fit single sloped model is inconsistent with the results from \citet{Bernstein2004} at $m(R)\gtrsim 28$ as the number of objects detected in that survey are too few by a factor of $\sim 6$ from that predicted by the model. To account for this, the LF must roll-over to shallower slopes at some faint magnitude not present in the F07 sample.

\subsection{The Size Distribution}

Our best-fit luminosity function slope is $\atotal$. Under the assumption that Eq.~\ref{eq:alpha-q} holds, we find that the KBO differential size distribution has a slope of $q=4.25$ with a 1-sigma uncertainty of $0.25$. If the belt is in a state of collisional equilibrium, we would expect the slope to be $q\simeq3.5$. This is inconsistent with the inferred size distribution at the 3-sigma level. We conclude that, for objects larger than $r \approx 50$ km, the size-distribution of the belt is inconsistent with a system in collisional equilibrium.

\section{Discussion}
To understand what the KBO size distribution tells us about the history of objects in the outer solar system, we must interpret our observations in terms of models that account for the size distribution of a belt of planetesimals. Generally, there are two broad types of models that attempt this, fragmentation models and accretion models. 

Analytic fragmentation models are those in which a series of equations accounting for accretion and collisional disruption are solved, producing a steady-state collisional-cascade equilibrium.  These models vary in complexity. Some assume that each disruption produces a number of equal size collision remnants \citep{Dohnanyi1969,Pan2005}, while some model object-disruptions with a distribution of collision remnant sizes \citep{Obrien2003}. Some calculations also include various models of KBO physical strengths \citep{Obrien2003}. The detailed results depend on the calculations, but generally the outcome  is a differential size distribution in a quasi-equilibrium steady-state with a slope equal to or less than Dohnanyi slope, $q\lesssim 3.5$. 

\citet{Pan2005}, assumed that the strengths of large Kuiper belt objects are gravity dominated and modeled a collisional cascade in which all collisions were purely disruptive (no accretion). They found that, for objects smaller than some break size, the size distribution slope was $q=3$ and could not  account for the steep slope observed for large KBOs with this model.

\citet{Obrien2003} accounted for accretion and fragmentation, and a variation of object strength as a function of size. They found that the equilibrium size distribution slope was related to the power-law slope describing the variation in object disruption energy per unit mass with size.  From this model, the Kuiper belt size distribution inferred from the F07 sample implies a unit disruption energy that decreases rapidly with increasing size. This is consistent with objects whose physical strength is dominated by tensile forces, but is inconsistent with objects whose strength is dominated by gravity; these objects have an increasing disruption energy with size. Objects in the tensile strength dominated regime are typically smaller than 1 km in size. The objects we observe here are likely to be gravity dominated and the scaling of KBO disruption energy versus size implied by the model from \citet{Obrien2003} seems unlikely. We conclude from these results that the observed LF and the inferred size distribution is inconsistent with analytic equilibrium models.

The large diversity of fragmentation models have similar outcomes; they predict a shallow size distribution slope of $q\sim 3.5$ which is incompatible with the large object size distribution inferred from the luminosity function of the Kuiper belt.

Numerical accretion models are those in which the size distribution, and orbits of a population of bodies is calculated. Unlike analytic collisional cascade models, the size distribution calculated from accretion models is not assumed to be in steady state, but evolves in time along with the orbital distribution. Models which account for accretion and fragmentation of planetesimals in the region of the Kuiper belt, predict a broken power-law size distribution with a steep slope  for large objects ($D \gtrsim 10$ km) that rolls over to a shallower slope for smaller objects ($D \sim 2$ km) \citep{Kenyon2001, Kenyon2002, Kenyon2004}. 

These models have two general evolutionary phases. The first phase is planetesimal accretion, in which a large reservoir of planetesimals on nearly circular orbits accrete to form larger bodies via low encounter-velocity collisions. This process rapidly produces a steep size distribution for large objects ($q\gtrsim 5$). Models from \citet{Kenyon2001} and \citet{Kenyon2002} calculate planet growth in the 40-47 AU zone. These models have a relatively low-mass intial Kuiper belt, and do not include stirring from Neptune. They  calculate planet growth before Neptune attains its current mass and orbit, and start from bodies with radius $\lesssim 1$km. Initially, the size distribution of large objects is very steep ($q \gtrsim 5$). After $\sim 100-300$ Myr, the slope of the largest objects flattens to $q\sim4.1-4.4$. 

The second phase is started when the belt members are stirred up onto dynamically excited orbits (via interactions with Neptune, oligarchs, etc.) such that mutual collisions are mainly catastrophic. Collisional grinding rapidly reduced the slope of the size distribution for small bodies ($D\lesssim 1$ km). The longer the collisional evolution continues, the larger the radius at which the break to shallower slopes occurs. In the models from \citet{Kenyon2001} and \citet{Kenyon2002}  the size distribution breaks to $q \lesssim 3.5$ at $D\sim2-20$km after $\sim1$ Gyr. 

More complicated models that start with a more massive initial Kuiper belt, and include effects of gas-drag and a more detailed model of object strength, produce large object slopes of $q \sim 3$ and after 4.5 Gyr, and a break at $D\sim 2-40$ km to very flat slopes of $q\sim 0-0.5$ for small objects.  \citep{Kenyon2004}

The observed size distribution exhibits a steeper slope than those determined by \citet{Kenyon2004} and is more consistent with models that start with a less massive Kuiper belt. In these models, the break radius grows larger and the large object slope flattens as accretion and collisional evolution continues. An accurate measure of the break size in addition to the large object slope determined here, combined with modeling, may further constrain the duration of accretion in the region of the Kuiper belt.

Accretion models which include migration of Neptune can account for some of the features of the KBO orbit distribution, but come at the cost of complicating the accretion process (Kenyon et al. 2007 and references there in). In these models, most of the current Kuiper belt consists of objects that are scattered by Neptune to their current positions, and originate from regions interior to Neptune's current orbit. Accretion of the KBOs in their original locations proceeds until stirring and scattering by Neptune disrupts the accretion process. 

In these migration models the KBOs are initially closer to the Sun than there current locations and accrete in regions of high density, and therefore both the accretion and collision process occur on much faster time-scales than in-situ formation models predict. The migration time-scale becomes an important parameter in these models; too fast a migration and the largest KBOs are not formed, while too slow, and the break size can evolve to be large enough such that it would be detected in the available observations.

Numerical accretion models of KBO formation, are generally in better agreement with the observed large size distribution than collisional cascade equilibrium models. Accretion models suggest that knowledge of the size distribution slope, and the radius at which the break to shallower slopes occurs will provide strong constraint on the accretion dynamics, and KBO formation time-scales. Due to the large degree of complexity in these models however, there is much work needed to interpret the observed size distribution. Models which achieve collisional-cascade equilibrium for large objects cannot account for the observed steep size distribution.

\section{Conclusions}
We have performed a survey of the Kuiper belt with an sky coverage of 3.0 square degrees with our deepest field having a depth of $m(g')=26.43$ magnitudes. An analysis of current survey data confirms that the luminosity function of the belt is well described by a single slope power-law between $m(R)=21-26$ with a slope of $\atotal$ and normalization $m_o=23.42 \pm 0.13$ which is typical of all the fields considered in the F07 sample.

We have shown the necessity of considering calibration, and density effects when inferring the luminosity function from the combination of different surveys that have observed separate regions of the sky, and are not directly calibrated to one-another. Such effects can skew the inferred LF away from the true form, and cause uncertainties in the inference to be highly underestimated. 

We conclude that:

\begin{enumerate}
	\item The Kuiper belt is not in a state of collisional equilibrium for objects larger than $D\approx 50$ km.

	\item There is no evidence for a change in the slope of the luminosity function for magnitudes brighter than $m(R)\sim 24.3$ corresponding to objects with diameters $D \gtrsim 110$ km at 40 AU. The sample of observations considered here is consistent with a best-fit model that breaks at $m(R)\sim24.4$ to a slope of $\alpha_2\sim0.6$, but the more complicated fit is not warranted by the slight improvement in the maximum likelihood value.

	\item The observed slope of the luminosity function implies that the size distribution of the Kuiper belt is consistent with a power-law with slope $q=4.25\pm0.25$ for objects with diameters larger than $D\sim50$km.

\section{Acknowledgements}
We thank the reviewers for providing excellent feedback and suggestions for improvement of this manuscript. We also thank Tom Loredo for his many discussions in the use of statistics.

This project was funded by the National Science and Engineering Research Council and the National Research Council of Canada. This research used the facilities of the Canadian Astronomy Data Centre operated by the National Research Council of Canada with the support of the Canadian Space Agency.

\appendix
\section{Psfmatch3 \label{ap:psfmatch3}}
Psfmatch3 is a subtraction routine that compensates
for variations between the image quality of the template image and that of 
individual images. A description of the basic algorithm follows.

Consider an observed image $O_{ij}$ ($i, j$ refer to pixel row and
column), and a suitably chosen reference image $I_{ij}$. We define an
$n_k \times n_k$ convolution kernel $K_{ij}$ such that the convolution
$K \ast I$ provides some ``best'' (in a least-squares sense)
estimator, $\tilde O$, of $O$; that is, the optimal kernel $K$ is the
one that minimizes $\sum_{i,j} (O_{ij}-{\tilde
  O}_{ij})^2$. Differentiating with respect to each kernel element
$K_{ij}$ yields a system of $M=n_k^2$ linear equations

{\tiny
\begin{equation}
\begin{array}{c}
	\left(
	\begin{array}{ccccc}
		\sum_{i,j}I_{i-1,j-1} I_{i-1,j-1} & \sum_{i,j}I_{i-2,j-1} I_{i-1,j-2} & ... & ... & ... \\
		... & \sum_{i,j}I_{i-2,j-2} I_{i-2,j-2} & ... & ... & ... \\
		... & ... & ... & ... & ... \\
		... & ... & ... & \sum_{i,j}I_{i-M+1,j-M+1} I_{i-M+1,j-M+1} & ... \\
		... & ... & ... & ... & \sum_{i,j}I_{i-M,j-M} I_{i-M,j-M} \\
	\end{array}
	\right) * \\
	
	\left(
	\begin{array}{c}
		K_1 \\
		K_2 \\
		... \\
		K_{M-1} \\
		K_{M} \\
	\end{array}
	\right) = \\
	\left(
	\begin{array}{c}
		\sum_{i,j} O_{ij} I_{i-1,j-1} \\
		... \\
		... \\
		... \\
		... \\
	\end{array}
	\right)\\
\label{eq:Chris_nasty}
\end{array}
\end{equation}
}

\noindent
which can be solved for the $n_k^2$ kernel elements.

This basic procedure can be improved considerably by adding an extra term for sky differences, viz.

  $$ \tilde O = K \ast I + \Delta s, $$

where $\Delta s$ is a constant denoting the difference in sky between the two images. Weighting by errors is easily included by minimizing $\sum_{i,j} {(O_{ij}-{\tilde O}_{ij})^2}\over{\sigma_{ij}^2}$. Perhaps most important is allowing for spatial variations in the kernel (and sky background); this is accomplished simply by solving for polynomial coefficients representing the spatial variation of each kernel coefficient.

There are a number of features and advantages to this method of matching the point spread function on two images. 
\begin{itemize}
	\item The kernel $K$ has arbitrary form; it does not need to be symmetric, and can handle PSF variations that may not always be handled by methods involving a basis set of functions to represent the kernel. The kernel can even be solved to perform deconvolution (which is the case when the reference image is erroneously chosen to have worse seeing), though this results in noise amplification. 
	\item $K$ is not necessarily normalized to unity; this automatically takes out transparency fluctuations. 
	\item The method automatically removes small shifts (even spatially variable shifts) between two images. 
	\item As noted above, the method can easily be adapted to include background differences, spatially variable backgrounds, and spatially variable kernels.
\end{itemize}

\end{enumerate}

\section{Field Divisions from \citet{Trujillo2001b}}

\begin{table}[h]
	\caption{Field Details. \label{atab:FieldData}}
		\begin{tabular}{llllll}
		Field & R.A. (hrs.)&Ecliptic Latitude ($^o$) & Area (Sq. $^o$) & \# Objects & Efficiency \\ \hline
		TE1G & 8 - 9 & -0.5 - 0.5 & 7.59 & 15 & Good \\
		TE1M & 8 - 9 & -0.5 - 0.5 & 3.3 & 4 & Medium \\
		TE2G & 10 - 11 & 0 & 4.35 & 11 & Good \\
		TE3G & 11 - 12.5 & 0 & 6.27 & 12 & Good \\
		TE3M & 11 - 12.5 & 0 & 2.31 & 4 & Medium \\
		TE4G & $\sim$21-23 & 0 & 2.64 & 4 & Good\\
		TE4M & $\sim$21-23 & 0 & 2.97 & 3 & Medium\\
		TE5G & $\sim$23 - 1 & 0 & 0.66 & 0 & Good \\
		TE5M & $\sim$23 - 1 & 0 & 4.39 & 7 & Medium 
		\end{tabular}
\end{table}

\newpage

\begin{table}[hp]
	\caption{\citet{Trujillo2001} pointings included in each field division. \label{atab:Pointings}}
		\begin{tabular}{l|l}
		Field & Pointings \\ \hline
		TE1G & 476727o, 476728o, 476729o, 476848o, 476849o, 476850o, 476851o, 476852o, 476853o,\\
		& 476854o, 476855o, 476856o, 476857o, 476858o, 476859o, 476984o, 476985o,\\
		& 476986o, 476987o, 476988o, 476989o, 476990o, 476991o\\ \hline
		TE1M & 476717o, 476718o, 476719o, 476720o, 476721o, 476992o, 476993o, 476994o, 476995o,\\ 
		& 476996o\\ \hline
		TE2G & 476885o, 476886o, 476887o, 476888o, 476889o, 476890o, 476891o, 476892o, 476893o,\\
		& 476894o, 476895o, 476896o,476924o\\ \hline
		TE3G & 476758o, 476759o, 476760o, 476761o, 476762o, 476763o, 476764o, 476765o, 476766o,\\ 
		& 476767o, 476768o, 476769o,476795o, 476796o, 476797o, 527174o, 527305o,\\
		& 527458o, 527461o\\  \hline
		TE3M & 476798o, 476799o, 527175o, 527306o, 527307o, 527459o, 527460o\\  \hline
		TE4G & 502047o, 502048o, 502049o, 502183o, 502184o, 502215o, 502217o, 502218o\\  \hline
		TE4M & 502050o, 502051o, 502052o, 502182o, 502185o, 502186o, 502214o, 502216o, 502374o\\  \hline
		TE5G & 502102o, 502139o\\  \hline
		TE5M & 502098o, 502099o, 502100o, 502101o, 502103o, 502136o, 502137o, 502138o, 502140o,\\
		& 502248o, 502249o, 502250o, 502426o\\  \hline
		\end{tabular}
\end{table}

\clearpage

%%%
%bibtex section
%%%
\bibliographystyle{I09874Style}
\bibliography{I09874bibliography}

\newpage

\begin{landscape}
\begin{table}[h]
   \caption{Field Details \label{tab:Fielddata}}
   {\footnotesize
	\begin{tabular}{lllcccccc}
         Field&Telescope&Date (UT)&Camera&Filter&Exp. Time&$\alpha$&$\delta$&Seeing (")\\ \hline
	UNE & CFHT & 2001-08-24 & CFH12k &Kron-Cousins R& 16 $\times$ 480 s & 21:41:06.6 & -14:28:04.9 & 1.0\\   
	UNW & CFHT & 2001-08-25 & CFH12k &Kron-Cousins R& 17 $\times$ 480 s & 21:38:34.9 & -14:37:15.6 & 0.87 \\
	MEGA & CFHT & 2004-09-15 & MEGAPrime & g' & 48 $\times$ 240 s & 22:24:46.4 & -10:46:51.3 & 0.74 \\
	& CFHT & 2004-09-16 &	 MEGAPrime & g' & 45 $\times$ 240 s & 22:24:37.9 & -10:47:39.0 & 0.83 \\
	N10032W3 & Blanco & 2001-08-10 & Mosaic2 & VR & 40 $\times$ 480 s & 20:38:35.8 &-18:01:02.5 & 1.4 \\
	N10033 & Blanco & 2001-08-11 & Mosaic2 & VR & 33 $\times$ 480 s & 20:39:05.8 & -18:37:45.5 & 1.3 \\
	NEP0813NW3 & Blanco & 2002-08-13 & Mosaic2 & VR & 21  $\times$ 480 s & 20:45:10 & -17:38:27.7 & 0.87 \\
	& Blanco & 2002-08-15 & Mosaic2 & VR & 20 $\times$ 480 s & 20:44:58.9 & -17:39:20.7 & 0.674\\
	NEP0815NE3 & Blanco & 2002-08-15 & Mosaic2 & VR & 21 $\times$ 480 s & 20:47:25.0 & -17:32:20.8 & 0.6\\
	\end{tabular}
	}
\end{table}
\end{landscape}

\newpage

\begin{table}[h]
   \caption{Observation Details. * - Field used only for confirmation purposes. $\lambda$ - ecliptic longitude $(^\circ)$. $\beta$ - ecliptic latitude $(^\circ)$. $\lambda_N$ - longitude with respect to Neptune $(^\circ)$. \label{tab:pointings}}
\begin{tabular}{lcccccccc}
	Field & Area $(^\circ square)$ & $\lambda$ &  $\beta$ & $\lambda_N$\\
	\hline
	UNE & 0.32 & 322.9 & -0.72 & 15.4\\
	UNW & 0.32 & 322.9 & -0.72 & 15.4 \\
	MEGAPrimeN1 &  0.85 & 334.9 & -0.76 & 19.9\\
	MEGAPrimeN2 * & - & 334.9 & -0.76 & 19.9\\
	N10032W3 & 0.38 & 307.4 & 0.13 & 0.0\\
	N11033 & 0.38 & 307.4 & 0.13 &  0.0\\
	NEP0813NW3 & 0.38 & 309.6 & 0.06 & 0.0 \\
	NEP0815NW3 * & - & 309.6 & 0.06 & 0.0 \\
	NEP0815NE3 & 0.38 & 309.6 & 0.06 & 0.0 \\
\end{tabular}   
\end{table}

\newpage

\begin{table}[h]
   \caption{Detection efficiency parameters using Eq.~\ref{eq:eff}, and flux measurement error parameter using Eq.~\ref{eq:dm_final}. A -  peak efficiency. $m_*$ - magnitude where efficiency is half the peak. g - width parameter. Z - field zeropoint. $\gamma$ - error parameter. \label{tab:effs}}
   \begin{tabular}{cccccc}
	Field & $\eta_{max}$  & $m_*$ & g & Z & $\gamma$\\
	\hline
	UNE & 0.96 & 25.32 (R) & 0.41 & 26.21 & 0.47 \\
	UNW & 0.97 & 25.44 (R) & 0.40 & 26.22 & 0.49 \\
	MEGAPrimeN1 &  0.97 & 26.43 (g') & 0.41 & 26.46 & 0.27 \\
	N10032W3 & 0.91 & 25.10 (VR) &  0.46 & 26.0 & 0.77 \\
	N11033 & 0.93 & 25.20 (VR) & 0.5 & 26.0 & 0.58\\
	NEP0813NW3 & 0.93 & 25.13 (VR) & 0.34 & 25.9 & 0.73\\
	NEP0815NE3 & 0.97 & 25.18 (VR) & 0.27 & 25.9 & 0.58\\
\end{tabular}
   
\end{table}

\newpage

\begin{longtable}{lcccc}

	% 72 objects, 50 used in the fits
	Object & m & Filter & Barycentric Distance (AU) & Inclination $(\circ)$\\
	\hline
	UNEa2 * & $23.23 \pm 0.03$ & R & $47 \pm 5$ & $  7 \pm 23$ \\ 
	UNEa4 * & $24.44 \pm 0.09$ & R & $45 \pm 6$ & $31 \pm 32$ \\ 
	UNEa6 * & $24.56 \pm 0.10$ & R & $48 \pm 5$ & $  2 \pm 22$ \\ 
	UNEa7 & $26.17 \pm 0.45$ & R & $49 \pm 5$ & $  0.6 \pm 8$ \\ 	\hline
	UNWa8 * & $23.17 \pm 0.07$ & R & $42 \pm 4$ & $  3 \pm 16$ \\ 	
	UNWb11 * & $24.59 \pm 0.11$ & R & $42 \pm 4$ & $  1 \pm 14$ \\ 	
	UNWa10 * & $25.05 \pm 0.17$ & R & $40 \pm 4$ & $  3 \pm 15$ \\ 	
	UNWa11 & $25.51 \pm 0.25$ & R & $44 \pm 4$ & $  11 \pm 19$ \\ 	
	UNWa6 & $25.61 \pm 0.28$ & R & $56 \pm 15$ & $  78 \pm 99$ \\ 	\hline
	MEGAa2 *C  & $24.10\pm 0.03$ & g' & $42 \pm 3$ & $18 \pm 7$ \\
	MEGAa33  *C & $24.60 \pm 0.05$ & g' & $43 \pm 3$ & $3 \pm 1$ \\
	MEGAb12  *C & $24.69 \pm 0.05$ & g' & $39 \pm 3$ & $25 \pm 11$ \\
	MEGAa10 *g & $25.25 \pm 0.09$ & g' & $41 \pm 5$ & $41 \pm 30$ \\
	MEGAb33  *CV & $25.25 \pm 0.09 (0.7)$ & g' & $44 \pm 3$ & $1.1 \pm 0.3$ \\
	MEGAa15  *C & $25.41 \pm 0.10$ & g' & $45 \pm 3$ & $1.0 \pm 1.0$ \\
	MEGAa23  *C & $25.59 \pm 0.12$ & g' & $41 \pm 3$ & $24 \pm 11$ \\
	MEGAa7 *C & $25.76 \pm 0.14$ & g' & $48 \pm 3$ & $1.3 \pm 1.3$ \\
	MEGAa24  *C & $25.86 \pm 0.16$ & g' & $47 \pm 3$ & $12 \pm 18$ \\
	MEGAa31  *CV & $25.86 \pm 0.16 (0.7)$ & g' & $40 \pm 3$ & $3 \pm 1$ \\
	MEGAc19  *C & $25.92 \pm 0.17$ & g' & $41 \pm 3$ & $1.2 \pm 0.5$ \\
	MEGAa29  *C & $25.93 \pm 0.17$ & g' & $41 \pm 3$ & $1.3 \pm 0.3$ \\
	MEGAa35  *C & $25.99 \pm 0.018$ & g' & $39 \pm 3$ & $30 \pm 14$ \\
	MEGAa19  *C & $26.07 \pm 0.19 $ & g' & $39 \pm 4$ & $38 \pm 20$ \\
	MEGAb29  *C & $26.16 \pm 0.21$ & g' & $39 \pm 5$ & $5 \pm 2$ \\
	MEGAa20  *C & $26.19 \pm 0.21$ & g' & $44 \pm 3$ & $2.1 \pm 1.1$ \\
	MEGAa22  *C & $26.24 \pm 0.22$ & g' & $45 \pm 3$ & $4 \pm 2$ \\
	MEGAa8 *C & $26.26 \pm 0.23$ & g' & $39 \pm 3$ & $8 \pm 3$ \\
	MEGAa14*g & $26.35 \pm 0.25$ & g' & $41 \pm 3$ & $4 \pm 12$ \\
	MEGAa5 *g & $26.39 \pm 0.25$ & g' & $37 \pm 3$ & $15 \pm 12$ \\
	MEGAa12*C & $26.42 \pm 0.26$ & g' & $42 \pm 3$ & $32 \pm 15$ \\
	MEGAa18 C & $26.60 \pm 0.31$ & g' & $41 \pm 3$ & $3 \pm 2$ \\
	MEGAb19 C & $26.62 \pm 0.31$ & g' & $64 \pm 7$ & $76 \pm 38$ \\
	MEGAa1  C  & $26.68 \pm 0.34$ & g' & $36 \pm 3$ & $12 \pm 5$ \\
	MEGAc33  & $27.03 \pm 0.45$ & g' & $42 \pm 4$ & $15 \pm 15$ \\   \hline
	N11033a5 * & $22.09  \pm 0.015$ & VR  & $36 \pm 4$ & $33 \pm 18$ \\
	N11033c1 * & $22.70 \pm 0.03$ & VR   & $46 \pm 3$ & $1 \pm 8$ \\
	N11033d13 * & $24.18 \pm 0.24$ & VR & $58 \pm 4$ & $2 \pm 12$ \\
	N11033b7 * & $24.61 \pm 0.16 $ & VR   & $43 \pm 3$ & $3 \pm 8$ \\
	N11033c7 * & $24.79 \pm 0.19$ & VR   & $44 \pm 3$ & $2 \pm 8$ \\
	N11033a9 * & $24.80 \pm 0.19$ & VR   & $46 \pm 3$ & $1 \pm 8$ \\
	N11033e1* & $25.19  \pm 0.27 $ & VR & $43 \pm 3$ & $4 \pm 7$ \\
	N11033c13* & $25.20 \pm 0.11$ & VR & $44 \pm 3$ & $20 \pm 12$ \\
	N11033b1* & $25.22 \pm 0.28$ & VR     & $44 \pm 4$ & $1 \pm 8$ \\
	N11033b9 & $25.282  \pm 0.30$ & VR & $45 \pm 3$ & $15 \pm 10$ \\
	N11033d1 & $25.44  \pm  0.34$ & VR & $43 \pm 3$ & $1 \pm 7$ \\
	N11033a11 & $25.50  \pm 0.37$ & VR & $43 \pm 3$ & $25 \pm 14$ \\
	N11033a13 & $25.51 \pm 0.28$ & VR & $45 \pm 3$ & $27 \pm 14$ \\
	N11033a1 & $25.94 \pm 0.55$ & VR   & $44\pm 3$ & $14 \pm 10$ \\   \hline
	N10032W3a5 * & $23.52 \pm 0.08$ & VR & $39 \pm 3$ & $5 \pm 6$ \\
	N10032W3c13 * & $23.54 \pm 0.08$ & VR & $45 \pm 3$ & $1 \pm 7$ \\
	N10032W3b13 * & $24.46 \pm 0.19$ & VR & $45 \pm 3$ & $4 \pm 7$ \\
	N10032W3a11 * & $24.52 \pm 0.20$ & VR & $44 \pm 3$ & $4 \pm 7$ \\
	N10032W3a15 * & $24.69 \pm 0.23$ & VR & $43 \pm 3$ & $1 \pm 6$ \\
	N10032W3b1 * & $24.97 \pm 0.29$ & VR & $45 \pm 3$ & $12 \pm 9$ \\
	N10032W3b15 * & $25.04 \pm 0.32 $ & VR & $44 \pm 3$ & $3 \pm 7$ \\
	N10032W3a3 * & $25.05 \pm 0.32$ & VR & $44 \pm 3$ & $1 \pm 7$ \\
	N10032W3a9 & $26.12 \pm 0.86$ & VR & $41 \pm 3$ & $17 \pm 10$ \\   \hline
	NEP0813NW3a5 *C & $23.58 \pm 0.06$ & VR &  $43 \pm 2.5$ & $ 1.4 \pm 0.6$\\
	NEP0813NW3a7 *C & $24.05 \pm 0.10$ & VR &  $45 \pm 2.6$ & $ 3 \pm 1$\\
	NEP0813NW3a1 *C & $24.62 \pm 0.20$ & VR &  $43 \pm 2.6$ & $ 0.8 \pm 0.6$\\
	NEP0813NW3b7 *C & $24.80 \pm 0.17$ & VR &  $42 \pm 2.5$ & $ 0.6 \pm 0.1$\\
	NEP0813NW3a11 C & $25.17 \pm 0.29$ & VR &  $45 \pm 2.7$ & $ 8 \pm 3$\\
	NEP0813NW3a3 C& $25.26 \pm 0.4$ & VR &  $39 \pm 2.7$ & $ 20 \pm 8$\\ \hline
	NEP0815NE3c9 * & $22.62 \pm 0.02$ & VR &  $41 \pm 3$ & $ 0.3 \pm 5$\\
	NEP0815NE3b15 * & $24.06 \pm 0.09$ & VR &  $43 \pm 3$ & $ 5 \pm 6$\\
	NEP0815NE3a3 * & $24.35 \pm 0.12$ & VR &  $39 \pm 3$ & $ 25 \pm 12$\\
	NEP0815NE3a15 * & $24.79 \pm 0.19$ & VR &  $44 \pm 3$ & $ 1 \pm 6$\\
	NEP0815NE3b9 * & $24.83 \pm 0.19$ & VR &  $47 \pm 3$ & $ 1 \pm 7$\\
	NEP0815NE3a13 * & $24.89 \pm 0.21$ & VR &  $43 \pm 3$ & $ 16 \pm 9$\\
	NEP0815NE3a1 * & $25.02 \pm 0.25$ & VR &  $46 \pm 3.1$ & $ 20 \pm 11$\\
	NEP0815NE3b13 * & $25.05 \pm 0.23$ & VR &  $42 \pm 3$ & $ 5 \pm 6$\\
	NEP0815NE3a5 & $25.49 \pm 0.36$ & VR &  $44 \pm 3$ & $ 6 \pm 6$\\
	\caption{Detections List. * - object used in likelihood fits. C - object confirmed on second night. g - object fell on chip gap on second night. V- object has variable magnitude. Number in brackets in column 2 is the range of magnitude measurements for the variable objects. \label{tab:detections}}   
\end{longtable}

\newpage

\begin{longtable}{lllllll}
	Field & Filter & N & $N_{50}$ & alpha & $m_o$ & Reference/Fields\\
	\hline
	UN & R & 9 & 6 &$0.34\pm0.26$ & $22.3\pm 2$ & UNE,UNW\\
	MEGA & g' & 25 & 21 & $0.76\pm0.25$ & $24.5\pm0.65$ & MEGA\\
	CTIO01 & VR & 23 & 14 & $0.56\pm0.25$ & $22.7\pm1.0$ & N10032W3, N11033\\
	CTIO02 & VR & 15 & 12 & $0.68\pm0.31$ & $23.3\pm1.0$ & NEP0813NW3, NEP0815NE3\\
	SSU & r & 38 & 20 & $0.94\pm0.31$ & $23.4\pm0.25$ & \citet{Petit2006} Uranus Fields \\
	SSN & r & 27 & 17 & $1.12\pm0.38$ & $23.3\pm0.3$ & \citet{Petit2006} Neptune Fields \\
	G01 & r & 17 & 14 & $0.5$* & $ 23$* & \citet{Gladman2001} CFHT Field \\
	AF & VR & 10 & 10 & $0.34\pm0.2$ & $21.4\pm2.0$ & \citet{Allen2001} AF Field\\
	AKL & VR & 8 & 7 & $0.78\pm0.45$ & $23.5\pm1.0$ & \citet{Allen2001} KL Fields\\
	TE1G & r & 15 & 12 & $0.56\pm0.25$ & $23.4\pm0.5$ & \citet{Trujillo2001b} See Appendix 2 \\
	TE2G & r & 7 & 6 & $0.44\pm0.3$ & $23.4\pm1.0$ & \citet{Trujillo2001b} See Appendix 2 \\
	TE3G & r & 12 & 7 & $0.74\pm0.45$ & $23.7\pm0.4$ & \citet{Trujillo2001b} See Appendix 2 \\
	TE5M & g' & 7 & 5 & $1.44\pm0.9$ & $23.4\pm1.5$ & \citet{Trujillo2001b} See Appendix 2 \\
	\caption{Results of the maximum likelihood fits for all fields fit individually. Uncertainties are taken from the extrema of the 1-sigma likelihood contours. \label{tab:results} *-Provides a 3-sigma lower limit only. N - Number of objects detected in the field. $N_{50}$ - number of objects brightward of 50 \% threshold.} 
\end{longtable}

\newpage

\section*{Fig. Captions}

\begin{figure}[hp] %  figure placement: here, top, bottom, or page
   \centering
   \includegraphics[width=6in]{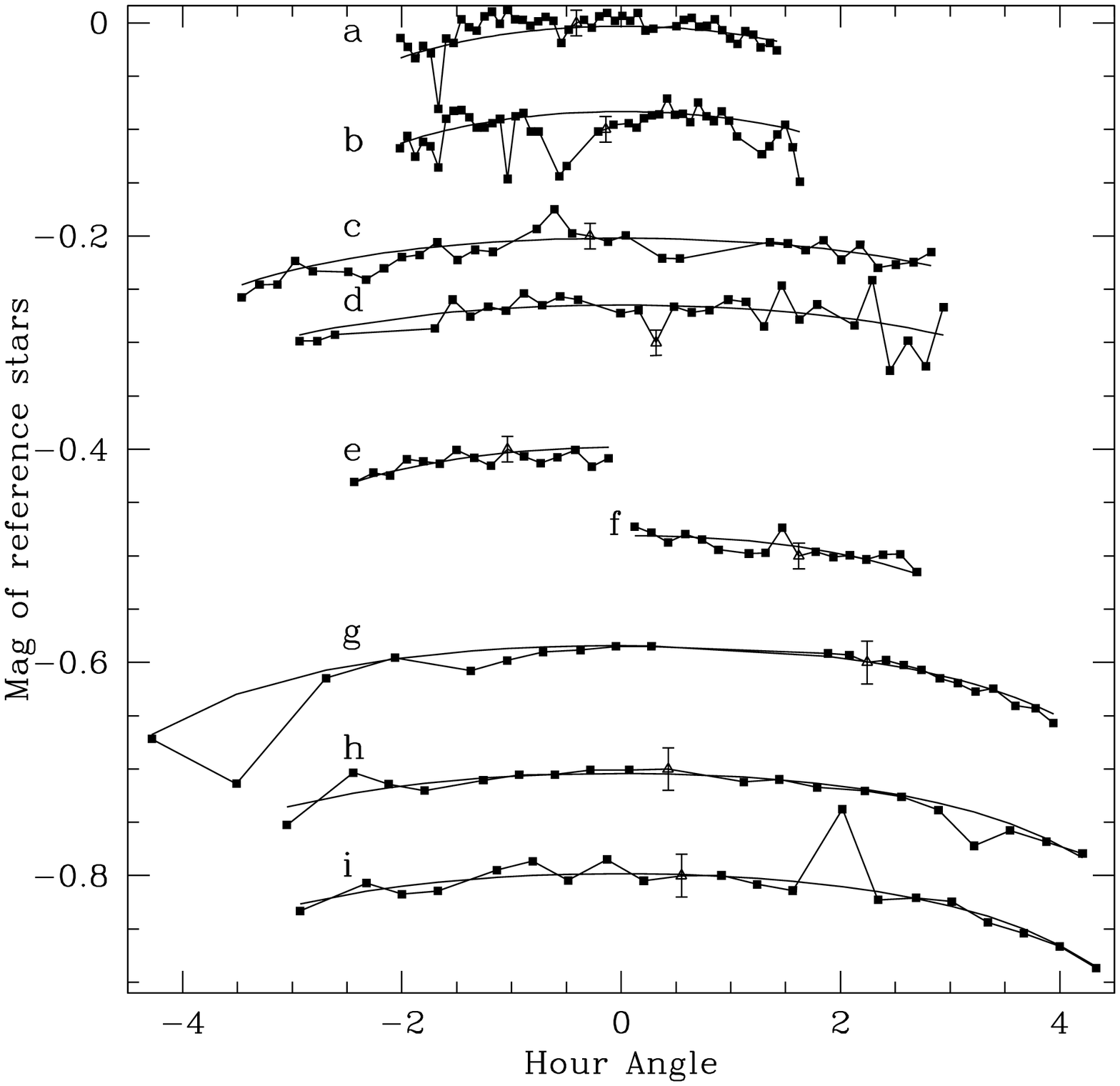}
   \caption{\label{fig:MAG_vs_HA}}
\end{figure}

Fig.~\ref{fig:MAG_vs_HA}~Variation of magnitude of reference stars versus hour angle of observations. Smooth line is the magnitude variation with airmass expected from nominal airmass extinction terms reported by the telescopes. The reference image used for planted object magnitude scaling is shown as an open triangle for each night, with a typical uncertainty shown for each point given. Each field is labeled; a: MEGAPrime Night1, b: MEGAPrime Night2, c: N10032W3, d: N11033, e: UNE, f: UNW, g: NEP0813NW3 Night1, h: NEP0815NW3 Night2, i: NEP0815NE3.

\begin{figure}[hp] %  figure placement: here, top, bottom, or page
   \centering
   \includegraphics[width=6in]{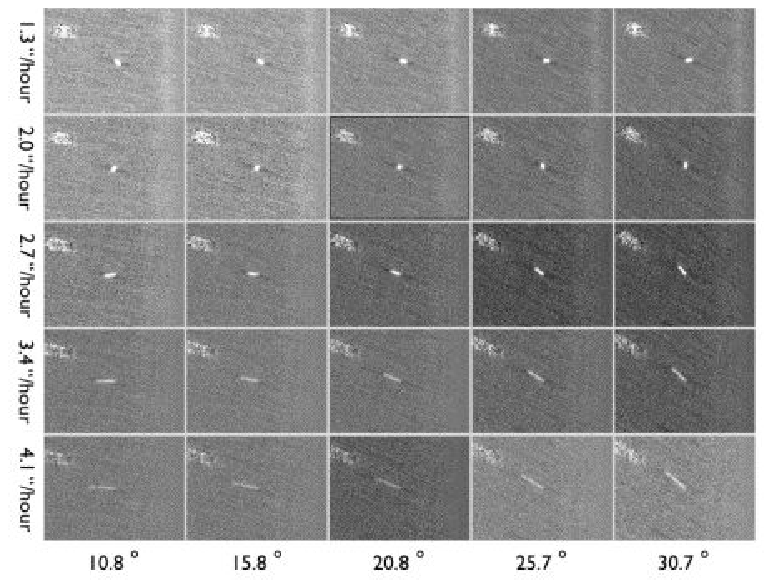} 
   \caption{\label{fig:shifts}}
\end{figure}

Fig.~\ref{fig:shifts}~Grid of shift rates and angles used to search the MEGAPrime field. Rates are in arcseconds per hour. Angles are in degrees below the horizontal. The source visible in the images is a real 23.8 mag. object with  a rate of motion close to 2 arcsec. hr$^{-1}$ at $20.8^\circ$, consistent with the ecliptic. The subtraction well and characteristic trailing can be seen at other rates of motion.

\begin{figure}[hp] %  figure placement: here, top, bottom, or page
   \centering
   \includegraphics[width=6in]{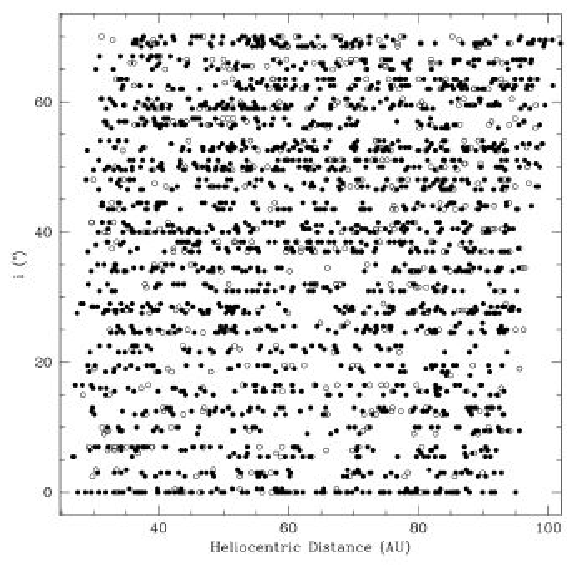} 
   \caption{\label{fig:r_i}}
\end{figure}

Fig.~\ref{fig:r_i}~Distribution of inclinations and distances for the artificial objects planted in the observations, taken from the planted objects lists of 10 chips in the MEGAPrime field. Solid circles mark those planted objects that were found during the object search. Open circles mark those that were not found. The same magnitude distribution of artificial objects was used at all distances. Thus, for a given magnitude, the detection efficiency did not depend on object distance.

\begin{figure}[hp] %  figure placement: here, top, bottom, or page
   \centering
   \includegraphics[width=6in]{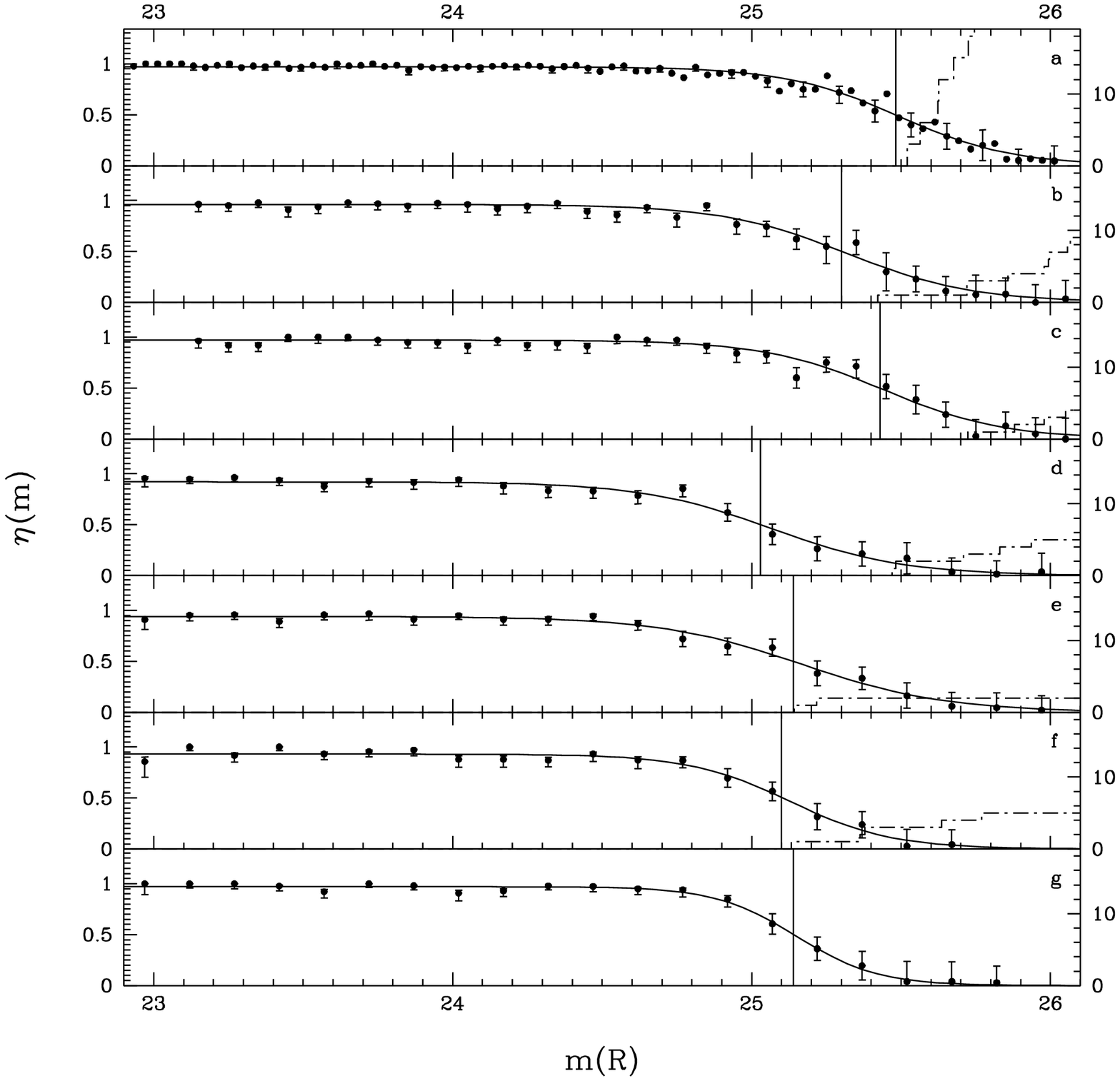} 
   \caption{\label{fig:eff}}
\end{figure}

Fig.~\ref{fig:eff}~Net efficiency for each field with 50\% efficiency marked with a vertical line. Magnitudes are converted to R-band using average colours (see Section~\ref{sec:Results}). Points are the binned efficiencies determined from the image search. Errorbars are 1-sigma Poisson confidence regions. The solid curve is the best fit efficiency using Eq.~\ref{eq:eff}. The dotted curve is a cumulative histogram of visually rejected false candidates. Each field is labeled; a: MEGAPrime N1, b: UNE, c: UNW, d: N10032W3, e: N11033,  f: NEP0813NW3, g: NEP0815NE3.

\begin{figure}[hp] %  figure placement: here, top, bottom, or page
   \centering
   \includegraphics[width=6in]{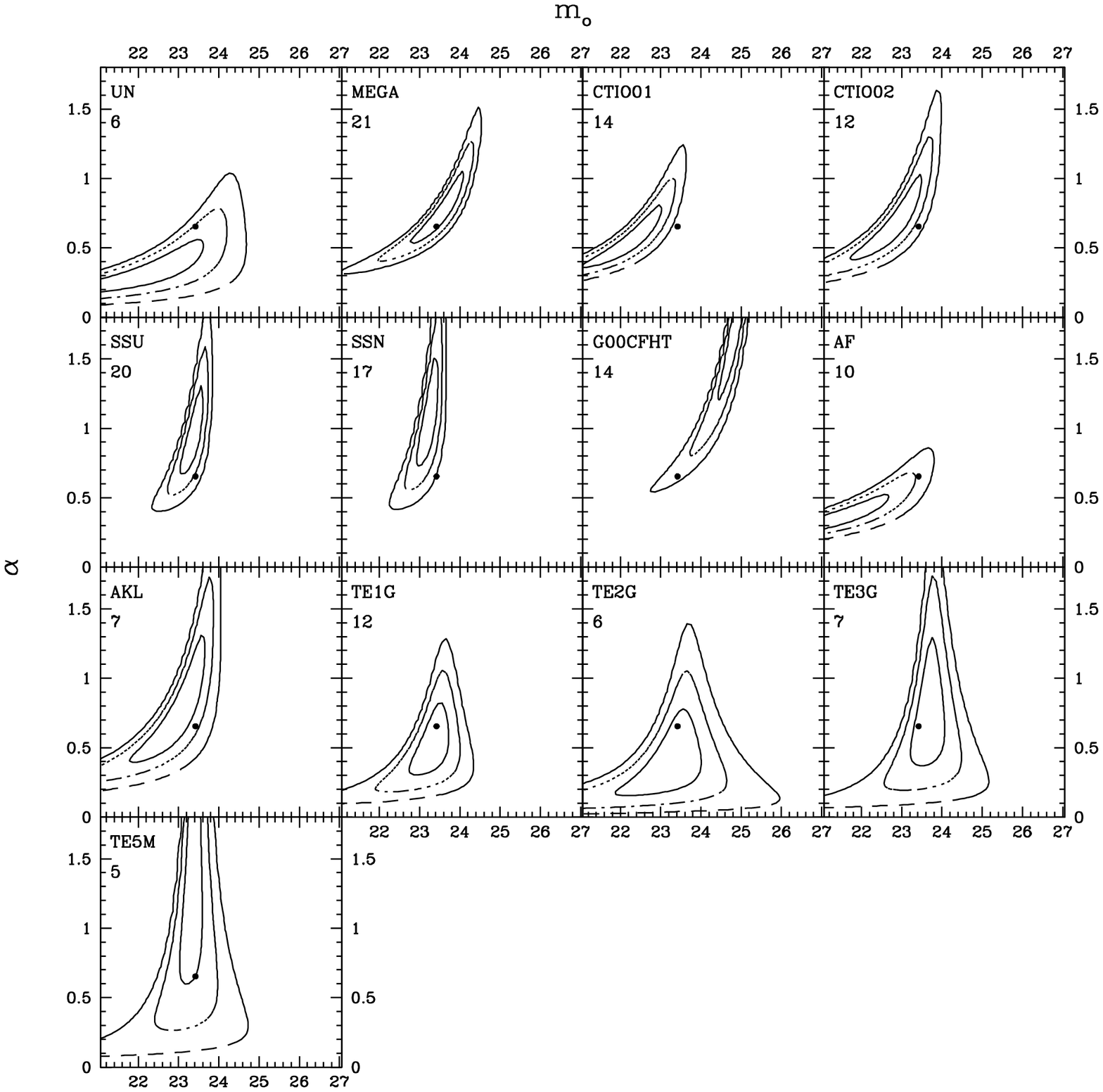} 
   \caption{\label{fig:likelihood}}
\end{figure}

Fig.~\ref{fig:likelihood}~Credible regions for the maximum likelihood fits to individual fields with 5 or more detections. Contours have been shifted to R-band using the nominal colours (see Section~\ref{sec:Results}). Solid: 1-sigma, Dashed: 2-sigma, Dotted: 3-sigma credibility contours. Field name and number of detections used in fits presented. Point indicates the best fit when combining all data together. Contours indicate $m_o$ is not consistent between surveys unless small shifts in $m_o$ are allowed.

\begin{figure}[hp] %  figure placement: here, top, bottom, or page
   \centering
   \includegraphics[width=6in]{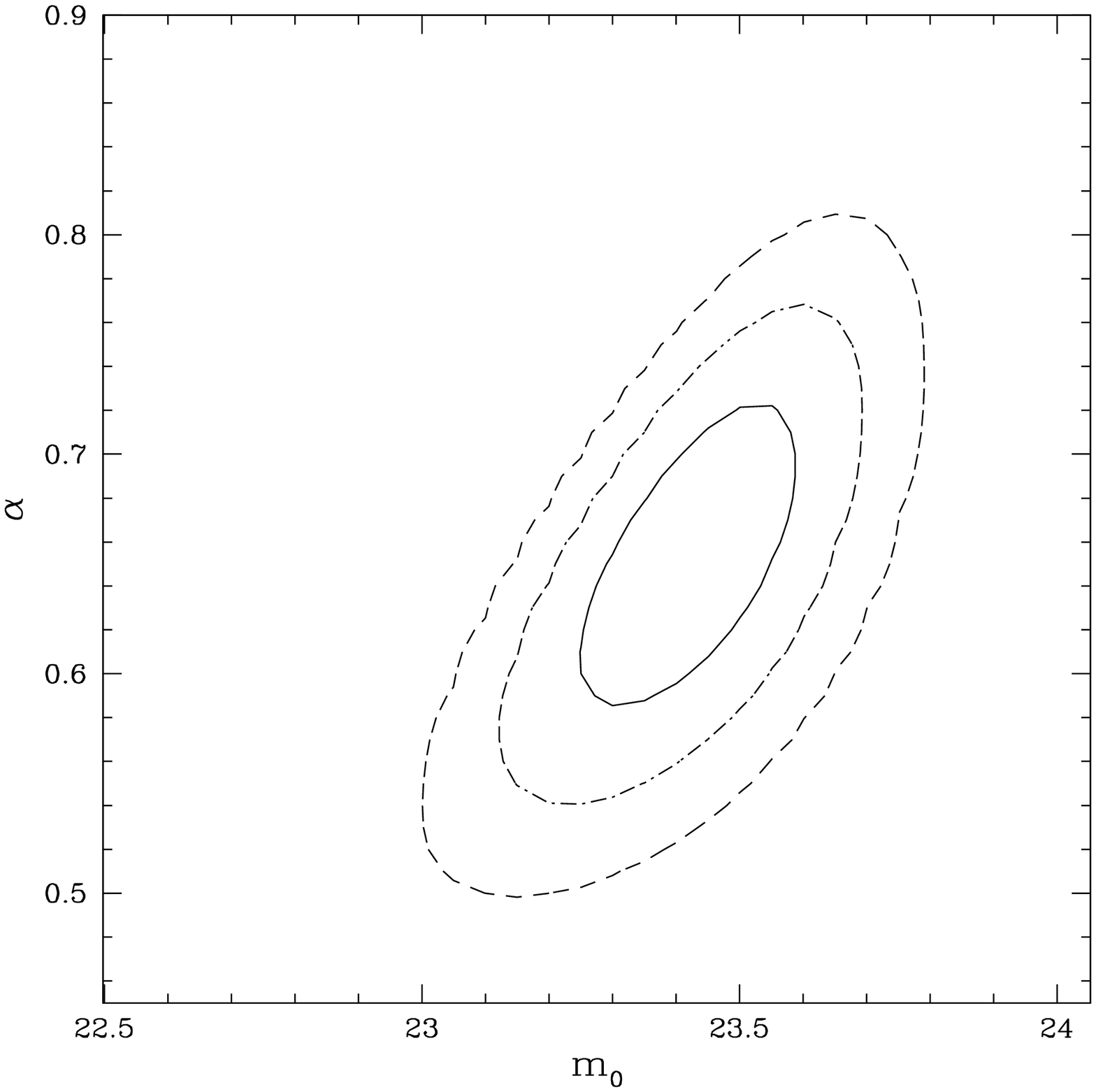} 
   \caption{\label{fig:likelihood_all}}
\end{figure}

Fig.~\ref{fig:likelihood_all}~1, 2, and 3-sigma confidence regions for the single power-law maximum likelihood fits using all data. Plotted are the likelihood contours for $\alpha$ and $m_{o}$.

\begin{figure}[hp] %  figure placement: here, top, bottom, or page
   \centering
   \includegraphics[width=6in]{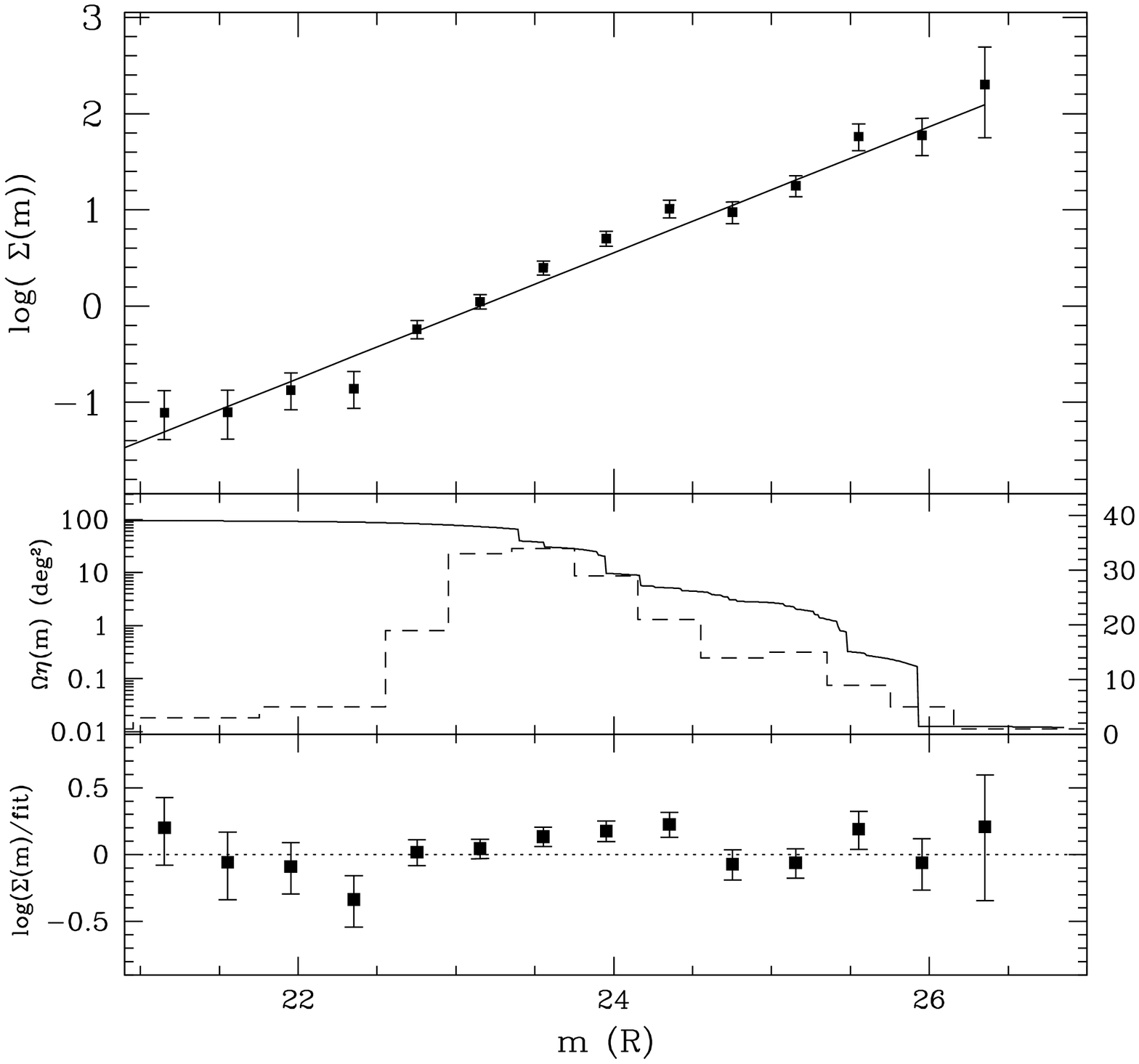} 
   \caption{\label{fig:cumul_diff}}
\end{figure}

Fig.~\ref{fig:cumul_diff}~{\it  Top: } Histogram of combined data using 0.4 mag bin-widths. Object magnitudes are shifted to R-band using typical magnitude colours. Errorbars are 1-sigma Poisson  intervals. Solid straight line: best fit line with $\alpha = 0.65$ and $m_o=23.42$. {\it Middle:} Effective area (area times efficiency) of combined data. Dashed line: histogram of all detections using 0.4 mag bin-width. {\it Bottom:} Log ratio of histogram to best fit $\alpha = 0.65$ and $m_o=23.42$. Errorbars are 1-sigma Poisson  interval.

\begin{figure}[hp] %  figure placement: here, top, bottom, or page
   \centering
   \includegraphics[width=6in]{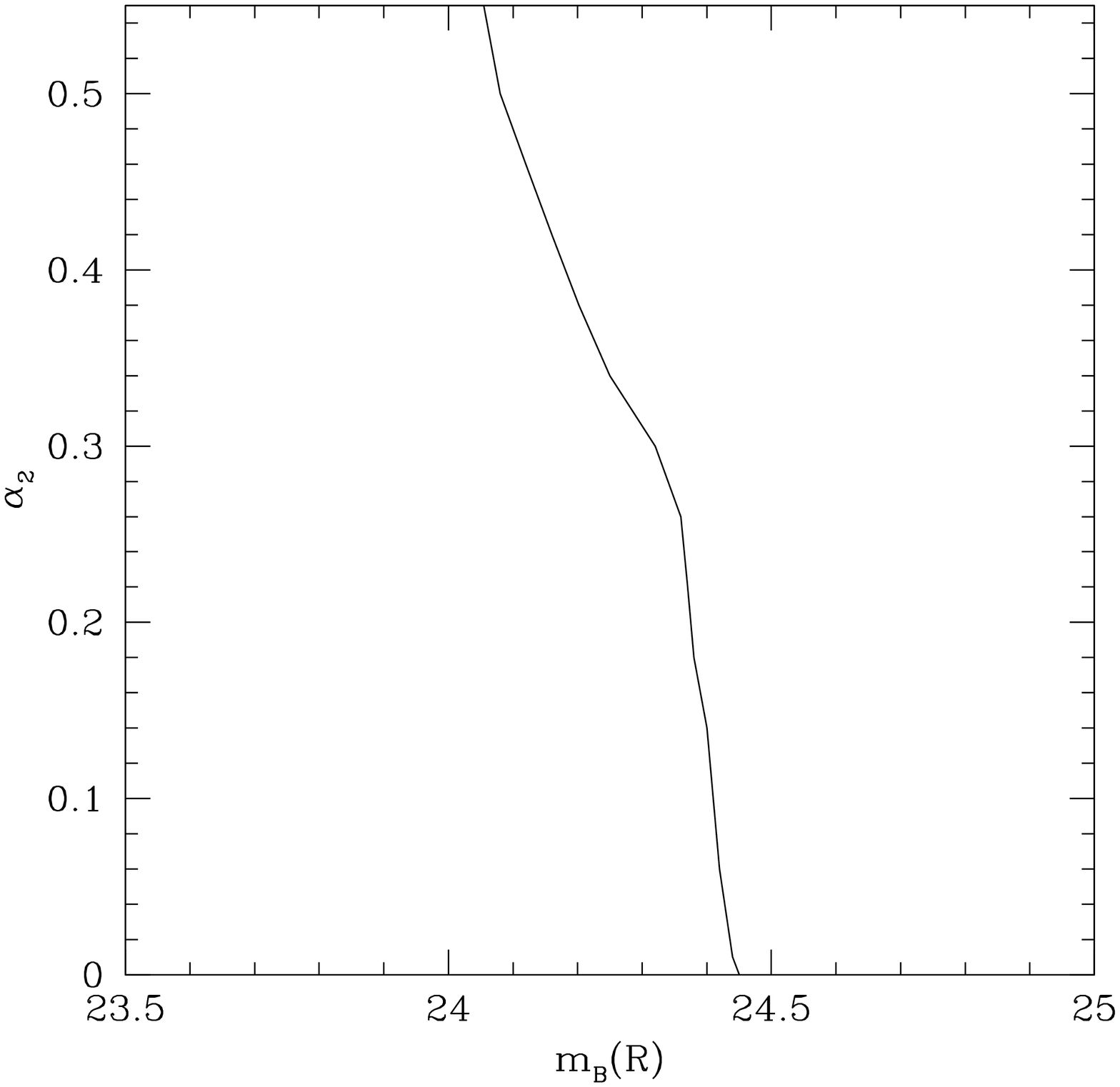} 
   \caption{\label{fig:KS}}
\end{figure}

Fig.~\ref{fig:KS}~Probability of rejection of a broken power-law from the data. Plotted is faint-end slope $\alpha_2$ versus break magnitude $m_B$. Contour represents the 95\% rejection confidence region. Contour has been smoothed to remove effects of course simulation sampling.

\newpage

\end{document}